\documentclass{elsart}
\usepackage{cite}
\usepackage{amssymb}
\usepackage{graphicx}

\newcommand{\ket}[1]{\left|  #1 \right\rangle}
\newcommand{\bracket}[3]{\langle {#1} | {#2} | {#3} \rangle}

\newcommand{\F}[0]{\ensuremath\mathcal{F}}
\newcommand{\FF}[0]{\ensuremath{F}}
\newcommand{\GG}[0]{\ensuremath{G}}
\newcommand{\aver}[1]{\ensuremath{\langle {#1} \rangle}}
\newcommand{\AverDist}[1]{\ensuremath{\{ {#1} \}}}
\newcommand{\im}[1]{\ensuremath{\mathrm{Im}(#1)}}
\newcommand{\re}[1]{\ensuremath{\mathrm{Re}(#1)}}

\newcommand{\Ein}[0]{\ensuremath{E_{in}}}
\newcommand{\Erad}[0]{\ensuremath{E_{rad}}}
\newcommand{\EM}[0]{\ensuremath{\mathcal{E}_\mathcal{M}}}
\newcommand{\EMN}[0]{\ensuremath{\mathcal{E}_{\mathcal{M}N}}}
\newcommand{\FM}[0]{\ensuremath{e_\mathcal{M}}}
\newcommand{\M}[0]{\ensuremath{\mathcal{M}}}
\newcommand{\Ec}[0]{\ensuremath{\mathcal{E}_c}}
\newcommand{\E}[0]{\ensuremath{\mathcal{E}}}
\newcommand{\EinMA}[0]{\ensuremath{\mathcal{E}_{in}}}

\newcommand{\PM}[0]{\ensuremath{P_\mathcal{M}}}
\newcommand{\PMN}[0]{\ensuremath{P_{\mathcal{M}N}}}
\newcommand{\Pfs}[0]{\ensuremath{P_{4\pi}}}
\newcommand{\PfsN}[0]{\ensuremath{P_{4\pi N}}}
\newcommand{\PcN}[0]{\ensuremath{P_{c N}}}
\newcommand{\Pzerofs}[0]{\ensuremath{P_{4\pi}^{(0)}}}
\newcommand{\Pabs}[0]{\ensuremath{P_{abs}}}
\newcommand{\Pin}[0]{\ensuremath{P_{in}}}
\newcommand{\Pc}[0]{\ensuremath{P_c}}
\newcommand{\Ptr}[0]{\ensuremath{P_{tr}}}
\newcommand{\Iin}[0]{\ensuremath{I_{in}}}
\newcommand{\Irad}[0]{\ensuremath{I_{rad}}}
\newcommand{\epsn}[0]{\ensuremath{\varepsilon_0}}
\newcommand{\etafs}[0]{\ensuremath{\eta_{fs}}}
\newcommand{\etac}[0]{\ensuremath{\eta}}
\newcommand{\dac}[0]{\ensuremath{\delta \omega_c}}
\newcommand{\dc}[0]{\ensuremath{\delta}}
\newcommand{\DA}[0]{\ensuremath{\Delta}}
\newcommand{\WA}[0]{\ensuremath{\omega_0}}
\newcommand{\wc}[0]{\ensuremath{\omega_c}}
\newcommand{\raw}[0]{\ensuremath{\rightarrow}}
\newcommand{\kM}[0]{\ensuremath{{\bf k}_\mathcal{M}}}
\newcommand{\La}[0]{\ensuremath{\mathcal{L}_a}}
\newcommand{\Ld}[0]{\ensuremath{\mathcal{L}_d}}

\begin{document}

\begin{frontmatter}



\title{Interaction between Atomic Ensembles and Optical Resonators: Classical Description}


\author[MIT,Harvard]{Haruka Tanji-Suzuki},
\author[MIT]{Ian D. Leroux},
\author[MIT]{Monika H. Schleier-Smith},
\author[MIT]{Marko Cetina},
\author[MIT]{Andrew T. Grier},
\author[Harvard]{Jonathan Simon}, and
\author[MIT]{Vladan Vuleti\'{c}}

\address[MIT]{Department of Physics,
MIT-Harvard Center for Ultracold Atoms, and Research Laboratory of
Electronics, Massachusetts Institute of Technology, Cambridge,
Massachusetts 02139, USA}
\address[Harvard]{Department of Physics, Harvard University, Cambridge,
Massachusetts 02138, USA}

\ead{vuletic@mit.edu}

\begin{abstract}
Many effects in the interaction between atoms and a cavity that are usually described in quantum mechanical terms (cavity quantum electrodynamics, cavity QED) can be understood and quantitatively analyzed within a classical framework.  We adopt such a classical picture of a radiating dipole oscillator to derive explicit expressions for the coupling of single atoms and atomic ensembles to Gaussian modes in free space and in an optical resonator. The cooperativity parameter of cavity QED is shown to play a central role, and is given a geometrical interpretation. The classical analysis yields transparent, intuitive results that are useful for analyzing applications of cavity QED such as atom detection and counting, cavity cooling, cavity spin squeezing, cavity spin optomechanics, or phase transitions associated with the self-organization of the ensemble-light system.

\end{abstract}


\end{frontmatter}

\section{Introduction}

The interaction of atoms with a single electromagnetic mode is a problem of significant fundamental interest. The quantum mechanical system consisting of a single atom interacting with a single mode can be analyzed exactly in the rotating-wave approximation for arbitrary coupling constant. This famous Jaynes-Cummings model \cite{Jaynes63} of cavity quantum electrodynamics (cavity QED) gives rise to many interesting effects such as Rabi oscillations with a single photon (vacuum Rabi oscillations), collapse and revival effects due to a dependence of the Rabi frequency on photon number, or optical nonlinearity at the single-photon level. Many of these effects have been observed in pioneering experiments both in the microwave domain by Haroche and coworkers \cite{Goy83,Kaluzny83,Haroche85} and Walther and coworkers \cite{Meschede85}, and in the optical domain by Kimble \cite{Rempe91,Thompson92,Turchette95,Kimble98,McKeever04,Birnbaum05,Boozer07}, Rempe \cite{Pinkse00,Kuhn02,Legero04,Maunz04,Nussmann05,Wilk07,Schuster08}, and others \cite{Heinzen87,Heinzen87a,Colombe07,Brennecke07}. Studies have concentrated on fundamental aspects of the system such as the vacuum Rabi splitting \cite{Kaluzny83,Agarwal84,Raizen89,Zhu90,Thompson92,Boca04,Colombe07,Brennecke07}, non-classical light generation \cite{Kuhn02,McKeever04a,Legero04,Thompson06,Simon07,Wilk07,Gupta07,Schuster08}, single-atom maser \cite{Meschede85} and laser operation \cite{McKeever03a}, or superradiance in the case of many atoms \cite{Raimond82,Kaluzny83,Slama07}. Significant effort has gone towards increasing the single-photon Rabi frequency $2g$ (also called vacuum Rabi frequency), at which a single quantum of excitation is exchanged between the atom and the cavity, above the dissipation rates $\kappa$ and $\Gamma$ at which the photon is lost from the cavity or from the atom by emission into free space, respectively. In this so-called strong-coupling limit of cavity QED, namely $2g \gg (\kappa, \Gamma)$, the coherent, reversible light-atom interaction dominates over dissipative processes. This should enable full quantum mechanical control over the atoms and photons, e.g., in the form of quantum gates between two atoms \cite{Pellizzari95} or quantum networks \cite{Cirac97}.

Besides being of fundamental interest, cavity QED enables an increasing number of applications related to atom detection \cite{McKeever04,Hope04,Hope05,Teper06,Puppe07,Trupke07,Poldy08,Heine09,Wilzbach09,Terraciano09,Gehr10,Bochmann10,Kohnen11} and manipulation - be it of the spatial degrees of freedom \cite{Muenstermann99,Hood00,Black05,Nussmann05a,Murch08} such as in cavity cooling \cite{Mossberg91,Cirac93,Cirac95,Horak97,Hechenblaikner98,Gangl00,Gangl00a,Vuletic00,Domokos01,McKeever03,Maunz04,Nussmann05,Murr06,Boozer06,Morigi07,Lev08,Leibrandt09}, feedback cooling \cite{Fischer02,Vuletic07,Koch10},  self-organization and the superradiant phase transition \cite{Black03,Black05,Baumann10,vonCube04,Slama07,Bux11}, or of the spin degrees of freedom such as in spin squeezing \cite{Kuzmich97,Kuzmich98,Kuzmich00,Genes03,Dantan03,Dantan03a,Appel09,Schleier-Smith10,Leroux10,Schleier-Smith10a,Leroux10a}, spin optomechanics \cite{Brahms10}, preparation of non-classical atomic states \cite{Mekhov07,Simon07a,Mekhov09,Mekhov09a}, or cavity-based quantum memories for light \cite{Black05a,Thompson06,Simon07,Simon07a,Tanji09}.

Many of the above applications make use of atomic ensembles rather than single atoms, in which case the complete quantum description of the ensemble-cavity interaction is non-trivial as it in general involves a very large Hilbert space \cite{Baragiola10}. (Under assumptions of symmetry, exact solutions are possible in a much smaller Hilbert space, see \cite{Tavis68}.) On the other hand, many of these applications operate via coherent (Rayleigh) scattering, while incoherent spontaneous emission \cite{Mollow69,API98} is either negligible or an undesired process whose effect can be estimated by means other than solving the problem exactly. In such circumstances, the full quantum description may not be necessary, and a simpler classical picture may yield the correct results and provide a complementary or more intuitive understanding. An example of this is cavity cooling, where the full quantum mechanical description yields complex dynamics \cite{Horak97,Zippilli05a}. However, in the relevant limit of interest for applications (large light-atom detuning and low saturation of the atomic transition) a classical model yields simple and correct results that can be understood in terms of cavity-enhanced coherent scattering \cite{Vuletic00,Vuletic01}.

Furthermore, it has become increasingly clear that features that were originally assigned a quantum mechanical origin, such as the vacuum Rabi splitting \cite{Agarwal84,API98}, can be in fact described within a classical framework, and arise simply from a combination of linear atomic absorption and dispersion \cite{Zhu90,Dowling93}. This is not surprising as in the limit of low saturation the atom can be modeled as a harmonic oscillator, and the classical theory of coupled harmonic oscillators (cavity mode and weakly driven atom) gives the same mode structure as the quantum mechanical treatment \cite{Auffeves08}. It can then be advantageous to use the classical theory - within its limits of applicability - to describe, and develop an intuition for, more complex problems involving atomic ensembles.

The classical description also leads to some results that are of course contained in the quantum theory, but that are not necessarily obvious within that formalism. For instance, the quantum description in terms of a vacuum Rabi frequency (that perhaps should be more appropriately called single-photon Rabi frequency) that scales inversely with the square root of the mode volume \cite{API98} may lead one to believe that strong coupling and coherent atom-light interaction require small cavity volume. However, the classical description immediately reveals that the mode area plays a more fundamental role than the mode volume. As discussed below, this feature is captured in the so-called cooperativity parameter $\etac = 4 g^2/(\kappa \Gamma)$ of cavity QED \cite{Kimble98}, that, as we shall show, is a geometric parameter that characterizes the absorptive, emissive, or dispersive coupling of an atom to the cavity mode.

In this work we will analyze the atom-cavity interaction from a classical point of view, and derive analytical formulas that remain valid quantum mechanically. We shall see that in this description the dimensionless cooperativity parameter $\etac$ governs all aspects of the atom-cavity interaction. A strong-coupling limit can be defined by the condition $\etac>1$, corresponding to a situation where we can no longer assume the atomic dipole to be driven by the unperturbed incident field, but have to self-consistently include the field emitted by the atom, and circulating in the cavity, into the total driving field. Thus for $\etac>1$ the back-action of the cavity field generated by the oscillating atomic dipole on that same dipole is not negligible. This leads, among other effects, to the interesting result known from a quantum mechanical analysis \cite{Alsing92} that the scattering by the atom into free space can be substantially modified by a cavity, even if the cavity subtends only a small solid angle.

For equal cavity and atomic linewidths, $\kappa=\Gamma$, the thus defined classical strong coupling condition $\etac>1$ is equivalent to the standard strong-coupling condition $2g>(\kappa,\Gamma)$ of cavity QED, but it is less stringent than the latter for $\kappa > \Gamma$ or $\kappa < \Gamma$. (The classical strong-coupling condition $\etac>1$ corresponds to the single-photon Rabi frequency $2g$ being larger that the geometric mean of the atomic and cavity linewidths.) In general, the system can be parameterized in terms of two dimensionless parameters, namely the ratios $g/\kappa$ and $g/\Gamma$ in the cavity QED description, or, in the classical description, the cooperativity parameter $\etac$ and the linewidth ratio $\kappa/\Gamma$ . The cavity QED strong-coupling condition $2g>(\kappa,\Gamma)$ corresponds to a normal-mode splitting that is much larger than the linewidths of the normal modes. In contrast, the less stringent classical condition $\etac>1$ also includes situations where the normal modes overlap within their linewidths, but destructive interference between them arises in a manner that is closely related to electromagnetically induced transparency \cite{Harris89,Harris97,Litvak02}.

In most cases the coherent emission into the cavity will be associated with the desired "signal" process, while the emission into free space constitutes a "noise" process that leads to atomic decoherence, motional heating etc. To understand the fundamental limitations to processes like cavity cooling, spin squeezing, spin optomechanics or phase transitions due to self organization, we must therefore quantify both the emission into the cavity mode of interest, and into all other (free-space) modes. In the following, we will usually express the results as power ratios that can be given simple physical or geometrical interpretations.

In the following, we always consider two different scenarios: In the ``scattering" or ``driven-atom" setup radiation is coupled into the mode of interest $\M$ via the atom that is driven by an external field incident from the side. In the "absorption/dispersion" or ``driven-mode" setup the mode of interest $\M$ is excited directly, and the atom modifies the field in $\M$ via forward scattering, while also emitting radiation into all other modes. We will analyze both scenarios for $\M$ being either a free-space mode or a cavity mode.

\section{Interaction between a single atom and a free-space mode}

In the following we analyze the interaction of a single atom, described as a pointlike classical dipole oscillator, with a single transverse electromagnetic mode in free space. We will consider a Gaussian TEM$_{00}$ mode with a waist $w$ that is at least somewhat larger than an optical wavelength $\lambda$, such that the paraxial approximation for the propagation of Gaussian beams \cite{Kogelnik66,SiegmanLasers} remains valid. The classical-oscillator description of the atom agrees with the quantum mechanical treatment in the limit where the saturation of the atomic transition is negligible, be it due to low beam intensity, or large detuning of the light from atomic resonances \cite{Mollow69,API98}. The assumption that the atom is pointlike, i.e., that it can be localized to a small fraction of an optical wavelength, implies that the atom's kinetic temperature is well above the recoil limit.

The electric-field component $\tilde{{\bf E}}(t)=\frac{1}{2}\hat{{\bf e}} E e^{-i \omega t} + c.c.$ of a linearly polarized light field oscillating at frequency $\omega=ck$ induces a proportional atomic dipole moment $\tilde{{\bf p}}=\frac{1}{2}\hat{{\bf e}} p e^{-i \omega t} + c.c.$ that is oscillating at the same frequency. Here $\hat{{\bf e}}$ is the unit polarization vector, and $p=\alpha E$ the amplitude of the induced dipole moment.
The complex polarizability $\alpha$ is given by (see, e.g., \cite{Grimm00,Milonni08})
\begin{equation}\label{Eq:Polarizability}
    \alpha = 6 \pi \epsn c^3 \frac{\Gamma/\WA^2}{\WA^2-\omega^2-i(\omega^3/\WA^2)\Gamma}.
\end{equation}
Here $\WA = c k_0 = 2 \pi c / \lambda_0$ denotes the atomic resonance frequency and $\Gamma$ is the linewidth of the atomic transition. Eq. \ref{Eq:Polarizability} is valid both classically and quantum mechanically. In the classical description, the oscillating electron is damped due to the emission of radiation, and $\Gamma=q^2 k^2/(6 \pi \epsn m c)$, where $m$ and $q$ are the electron charge and mass, respectively (see, e.g., \cite{Berman08}). In the quantum mechanical description, $\Gamma= k_0^3 |\mu|^2/(3 \pi \epsilon_0 \hbar)$ is the spontaneous population decay rate of the atomic excited state, given in terms of the dipole matrix element $\mu \equiv \bracket{e}{\mu}{g}$ between ground state $\ket{g}$ and excited state $\ket{e}$. Due to the validity of Eq. \ref{Eq:Polarizability} in both the classical and quantum domains, the classical results we will derive below agree with the semiclassical results derived from quantum theory in the limit of low saturation of the atomic transition.

The polarizability $\alpha$ obeys the relation
\begin{equation}\label{Eq:PolarizabilityRelation}
    |\alpha|^2 = \frac{6 \pi \epsn}{k^3} \im{\alpha},
\end{equation}
which will be useful in relating the total scattered power, proportional to $|\alpha|^2$, to the absorption, given by the out-of-phase component of the forward-scattered field that is proportional to $\im{\alpha}$ (see Section \ref{Sec:AbsorptionFreeSpace}). Eq. \ref{Eq:PolarizabilityRelation} ensures that the optical theorem is satisfied, i.e., that the rate at which energy is absorbed from the incident mode by the atom equals the power scattered into other field modes \cite{Berman06,Berman08}.

The oscillating dipole emits a radiation field whose amplitude at large distance $R \gg \lambda$ from the atom is given by \cite{JacksonCED}
\begin{equation}\label{Eq:RadiatedField}
    \Erad(R,\theta)= \frac{k^2 \sin \theta}{4 \pi \epsn} \frac{e^{ikR}}{R} \alpha E,
\end{equation}
where $\theta$ is the angle between the polarization $\hat{{\bf e}}$ of the driving field and the direction of observation.

A fraction of the radiated power can be collected in some mode of interest. The field radiated into the same mode as the driving field can interfere with the latter, resulting in attenuation of the driving field, i.e., absorption, and a phase shift of the total field, i.e., dispersion. In the following sections, we derive simple expressions for these quantities, and interpret them geometrically.

\subsection{Scattering into a free-space mode: emission}\label{Sec:ScatteringFreeSpace}

\begin{figure}
\centering{\includegraphics[width=3.5in]{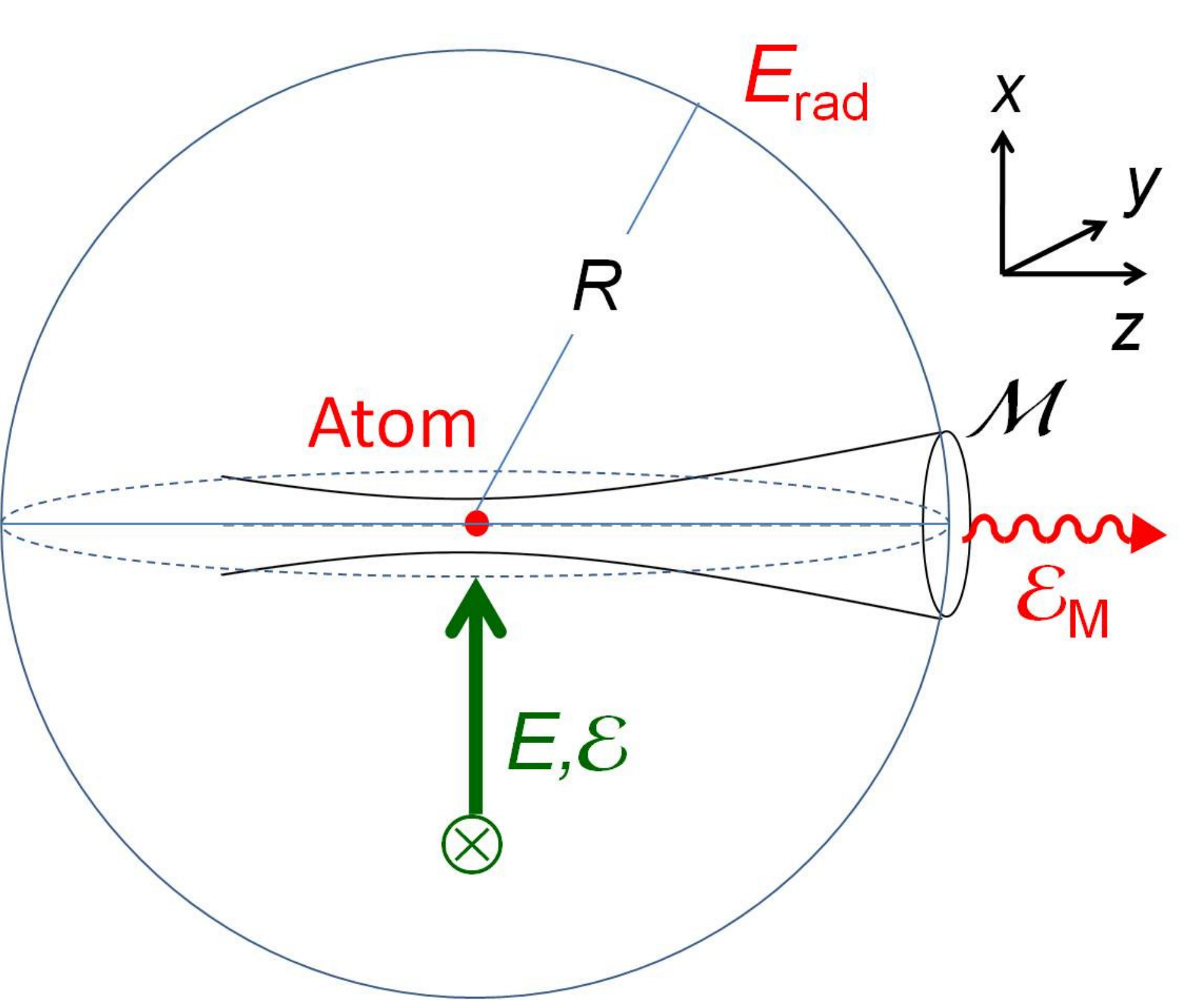}}
\caption{\label{Fig:FSScatteringSetup} Scattering of radiation by a weakly driven atom. The incident field $\E$ is polarized perpendicular to the TEM$_{00}$ mode of interest $\M$ and drives an atomic dipole oscillator that emits an electromagnetic field $\Erad$ at large distance $R$ from the atom. For the analysis we choose $R$ much larger than the Rayleigh range $z_R$ of $\M$.}
\end{figure}

We consider a traveling-wave TEM$_{00}$ Gaussian mode $\M$ of wavenumber  $k=2 \pi / \lambda = \omega/c$, waist $w$, and Rayleigh range $z_R=\pi w^2 / \lambda$. The atom is located on the axis of that mode at the waist, as shown in Fig. \ref{Fig:FSScatteringSetup}, and driven by an external field $E$ propagating in some other direction. The driving field polarization is assumed to be linear and perpendicular to the direction of propagation of the mode $\M$. We would like to know what fraction of the total power scattered by the driven atom is emitted into $\M$. This question can be answered by decomposing the dipole emission pattern into Hermite-Gaussian modes in a tangential plane located at distance $z=R \gg z_R$ in the far field (see Fig. \ref{Fig:FSScatteringSetup}). The normalized mode function $\FM(\rho,z)$ can be found in \cite{SiegmanLasers}, and in the tangential plane at $z \gg z_R$ is approximately
\begin{equation}\label{Eq:ModeFunction}
    \FM(\rho,z) \approx \left( \frac{2}{\pi \tilde{w}^2} \right)^{1/2} \exp\left( -\frac{\rho^2}{\tilde{w}^2}+ ikz + ik \frac{\rho^2}{2z}- i\frac{\pi}{2} \right).
\end{equation}
Here the first term in the exponent accounts for the intensity profile of the expanding Gaussian beam with $\tilde{w}(z)=w\sqrt{1+\left( z/z_R \right)^2} \approx w z/z_R$, The second and third term describe the beam wavefronts, and the last term is the Gouy phase shift of $\pi/2$ at $z \gg z_R$.

In general, the electric field $E_{\M}(\rho,z)$ in mode $\M$ at position $(\rho,z)$ can be written as $E_{\M}(\rho,z)=\FM(\rho,z)\EM /\sqrt{\epsn c}$ in terms of a position-independent quantity $\EM$ that we will refer to as the mode amplitude. $\EM$ is related to the total power $\PM$ in mode $\M$ via $\PM=|\EM|^2/2$, and to the electric field at the waist $E_{\M}(0,0)$ via $\EM=E_{\M}(0,0)\sqrt{\epsn c A}$, where $A = \pi w^2/2$ is the effective mode area. In the following it will be useful to similarly formally define a mode amplitude for the field $E$ driving the atom as $\E = \sqrt{\epsn c A} E$, even if the driving field is some arbitrary mode. As the induced dipole depends only on the electric field $E$ at the atom's position, all atomic absorption and emission can be expressed in terms of the rescaled quantity $\E$.

The mode $\M$ with $w \gg \lambda$ subtends only a small far-field angle $\lambda/(\pi w) \ll 1$ \cite{SiegmanLasers}, such that the spatial dependence of the emitted dipole field $\Erad$, Eq. \ref{Eq:RadiatedField}, over the region occupied by $\M$ can be approximated as $\sin \theta \approx 1$ and $e^{ikR}/R \approx e^{ikz + ik \rho^2/(2z)}/z$. Then the the mode amplitude $\EM$ arising from the radiated field can be calculated easily as the projection $\EM =  \sqrt{\epsn c} \int \FM^\ast \Erad 2\pi \rho d\rho$ in the plane at $z \gg z_R$. This yields the simple result
\begin{equation}\label{Eq:EMiBetaE}
     \EM= i \beta \E
\end{equation}
in terms of a dimensionless parameter
\begin{equation}\label{Eq:Beta}
    \beta =\frac{k}{\pi w^2}  \frac{\alpha}{\epsn}
\end{equation}
that characterizes the coupling of the incident field $\E$ to mode $\M$ via the atom with polarizability $\alpha$ at the drive frequency $ck$. From  Eq. \ref{Eq:PolarizabilityRelation} it follows that $\beta$ obeys the optical-theorem relation
\begin{equation}\label{Eq:BetaRelation}
    |\beta|^2 = \frac{6}{k^2 w^2} \im{\beta}=\etafs \im{\beta},
\end{equation}
where we have defined another dimensionless parameter, which we will call the single-atom cooperativity in free space, as
\begin{equation}\label{Eq:FSEta}
    \etafs=\frac{6}{k^2 w^2}.
\end{equation}

The total scattered power into all directions $\Pfs$ can be calculated by integrating the intensity $\Irad=\epsn c |\Erad|^2/2$ of the radiated field, Eq. \ref{Eq:RadiatedField}, over the surface of the sphere of radius $R$. Using Eqs. \ref{Eq:EMiBetaE}, \ref{Eq:Beta}, \ref{Eq:BetaRelation} the total emitted power can be expressed as
 \begin{equation}\label{Eq:FSPower4Pi}
    \Pfs= \frac{c k^4}{12 \pi \epsn} |\alpha E|^2= \im{\beta} |\E|^2 =  \frac{1}{\etafs} |\EM|^2.
\end{equation}

The power emitted into both directions of mode $\M $ is $2\PM= |\EM|^2$, and hence the cooperativity $\etafs$ is equal to the ratio of (bidirectional) emission into mode $\M$ and free-space emission $\Pfs$,
\begin{equation}\label{Eq:FSMFreeSpacePowerRatio}
    \frac{2\PM}{\Pfs} =\etafs,
\end{equation}
independent of the light frequency or value of the atomic polarizability. The free-space cooperativity $\etafs$ is a purely geometric quantity, and can be interpreted as the mode of interest subtending (bidirectionally) the effective solid angle $\Delta \Omega = 4/(k^2 w^2)$. An additional factor $3/2$ accounts for the directionality of the dipole emission pattern, and would be absent if the atomic dipole was driven by unpolarized light. Eq. \ref{Eq:FSMFreeSpacePowerRatio} is correct to lowest order in $(kw)^{-2} \ll 1$, and thus valid as long as the mode of interest is not focussed too strongly, i.e., $w \gtrsim \lambda$.

\subsection{Scattering from a free-space mode: absorption} \label{Sec:AbsorptionFreeSpace}

\begin{figure}
\centering{\includegraphics[width=5in]{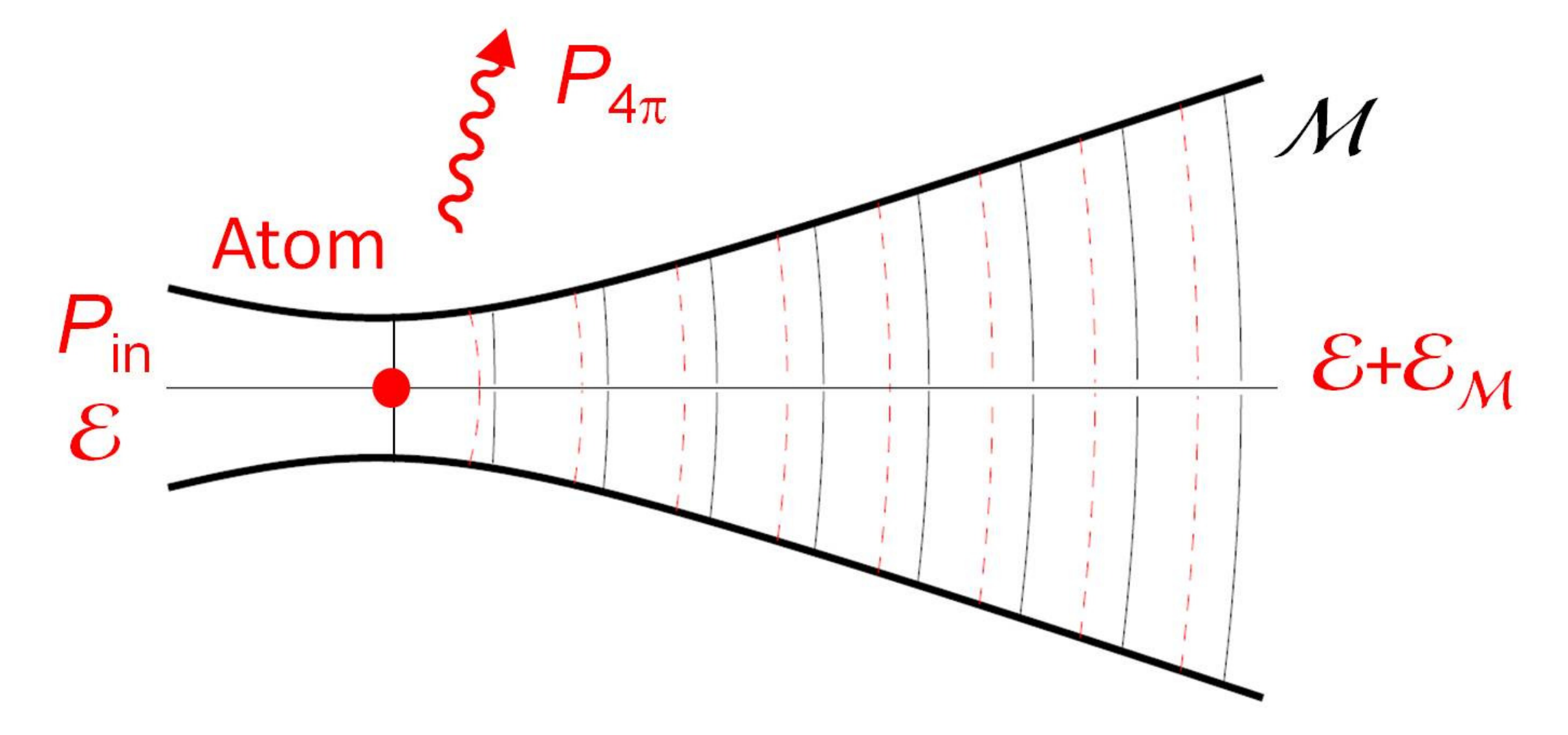}}
\caption{\label{Fig:FSAbsorptionSetup} Absorption by an atom placed at the center of a TEM$_{00}$ mode $\M$. The absorption can be calculated from the power $\Pfs$ radiated into free space, or from the field $\EM$ emitted by the atom in the forward direction that interferes with the incident field $\E$.}
\end{figure}

We consider the same mode $\M$ as in the previous section \ref{Sec:ScatteringFreeSpace}, but now take the light to be incident in that mode with power $\Pin=|\E|^2/2$, as shown in Fig. \ref{Fig:FSAbsorptionSetup}. The power $\Pfs$ scattered by the atom located at the mode waist on the mode axis, as given by Eq. \ref{Eq:FSPower4Pi}, by virtue of energy conservation must equal the power $\Pabs$ absorbed from the driving field. Then the fractional attenuation can be expressed as
\begin{equation}\label{Eq:FSAbsorption}
    \frac{\Pabs}{\Pin} = \frac{\Pfs}{\Pin}=\im{2\beta}.
\end{equation}

Within the rotating wave approximation (RWA), $\DA \equiv \omega-\WA \ll \WA$, the mode coupling parameter $\beta$ in terms of the light-atom detuning $\DA$ takes the simple form
\begin{equation}\label{Eq:BetaRWA}
    \beta_{RWA} = \etafs \left( \Ld(\DA) + i \La(\DA) \right),
\end{equation}
where $\La(\DA)=\Gamma^2/(\Gamma^2+4 \Delta^2)$ and $\Ld(\DA)=-2 \DA \Gamma /(\Gamma^2+4 \Delta^2)$ are the Lorentzian absorptive and dispersive lineshapes, respectively. Then the fractional attenuation can be written as
\begin{equation}\label{Eq:FSAbsorptionRWA}
    \left( \frac{\Pfs}{\Pin} \right)_{RWA}= 2 \etafs \La(\DA).
\end{equation}
On resonance ($\DA=0$) the beam attenuation (single-atom optical depth) equals twice the free-space cooperativity $\etafs$. These results are valid for $w \gtrsim \lambda$, i.e. for $\etafs \lesssim 6/(2\pi)^2 \approx 0.2$. Comparison of Eqs. \ref{Eq:FSMFreeSpacePowerRatio},\ref{Eq:FSAbsorptionRWA} reveals that the same geometric parameter $\etafs$ governs the fractional emission by the atom into a particular mode, and the resonant fractional absorption from a mode of the same geometry.

The atomic scattering cross section $\sigma$ is defined as the ratio of scattered power $\Pfs$ and incident intensity $\Iin=\Pin/A$,
\begin{equation}\label{Eq:CrossSection}
    \sigma= \frac{\Pfs}{\Iin} = \im{2\beta} A.
\end{equation}
In the RWA the scattering cross section according to Eq. \ref{Eq:BetaRWA} is given by
\begin{equation}\label{Eq:CrossSectionRWA}
    \sigma= \frac{6 \pi}{k_0^2} \La(\DA).
\end{equation}
The resonant absorption, and hence the cooperativity $\etafs=6/(k^2w^2) \approx 6/(k_0^2 w^2)$, can thus also be understood in terms of the ratio of the resonant scattering cross section $\sigma_0=6\pi/k_0^2$ and effective beam area $A=\pi w^2/2$, i.e. $\etafs \approx \sigma_0/(2A)$.

It is instructive to derive the atomic absorption from the requirement that the power reduction in the forward direction must be arising from destructive interference between the incident field $\E$ and the field $\EM=i\beta \E$ (Eq. \ref{Eq:EMiBetaE}) forward-scattered by the atom into the same mode $\M$. The total mode amplitude in the forward direction is $\E+\EM$, and the fractional absorption can be calculated as
\begin{equation}\label{Eq:FractionalAbsorption}
    \frac{\Pabs}{\Pin}= \frac{|\E|^2-|\E+\EM|^2}{|\E|^2} \approx -\frac{\E\EM^*+\E^* \EM}{|\E|^2} = \im{2 \beta},
\end{equation}
in agreement with the derivation based on the the radiated power $\Pfs$ (Eq. \ref{Eq:FSAbsorption}). In Eq. \ref{Eq:FractionalAbsorption} we have neglected the term $|\EM|^2$ that is smaller by a factor $(k w)^{-2} \ll 1$.

Note that the polarizability $\alpha$ on resonance is purely imaginary. Therefore from the expression for the radiated field $\Erad$, Eq. \ref{Eq:RadiatedField}, it would appear that the forward-scattered field on resonance is $\pi/2$ out of phase with the incident field, and thus cannot cancel the latter. However, we must keep in mind that the field $\Erad$ in Eq. \ref{Eq:RadiatedField} is a radial wave, while the input field is a Gaussian mode. To understand the implication of this, we can decompose the radial wave into Gaussian modes, or equivalently, consider the relative phase in the far field, where both modes are approximately spherical waves, and therefore interfere directly. In the far field $z \gg z_R$ there is a $\pi/2$ Gouy phase shift of the input field (relative to the driving field at the waist) \cite{Kogelnik66,SiegmanLasers}, as obvious from the mode function, Eq. \ref{Eq:ModeFunction}, and indicated by the wavefronts in Fig. \ref{Fig:FSAbsorptionSetup}. This additional phase shift of $\pi/2$ ensures that on atomic resonance the forward-scattered field destructively interferes with the input field. The above derivation represents a version of the optical theorem that states that the total scattered power $\Pfs$ is proportional to the imaginary part of the forward-scattering amplitude (see, e.g.,  \cite{Berman06,JacksonCED}).

\subsection{Phase shift of a free-space mode: dispersion} \label{Sec:FSDipsersion}

In general, the driving field in mode $\M$ is not only attenuated, but also experiences a phase shift in the presence of the atom. This phase shift, corresponding to the atomic index of refraction, can be simply understood as arising from the interference of the out-of-phase component of the forward-scattered field by the atom $\EM$ with the incident field in the same mode $\E$ \cite{FeynmanLecturesI}. Writing the field in the forward direction using Eq. \ref{Eq:EMiBetaE} as $\E + \EM = (1+i\beta)\E \approx \e^{i\beta}\E$, we see that the atom-induced phase shift of the light is
\begin{equation}\label{Eq:PhaseShiftFreeSpace}
    \phi= \re{\beta}.
\end{equation}

In the RWA the atom-induced phase shift of the incident mode for $\DA \gg \Gamma$ can  be written as
\begin{equation}\label{Eq:PhaseShiftFreeSpaceEta}
    \phi_{RWA}=\etafs \Ld(\DA) \approx -\etafs \frac{\Gamma}{2\DA}.
\end{equation}
At large detuning $\DA \gg \Gamma$ from atomic resonance the real part of the polarizability exceeds the imaginary part by a factor $\DA/\Gamma$, so the dispersion dominates the absorption (Eq. \ref{Eq:BetaRWA}).We see that the effect of the atom's index of refraction on the Gaussian mode also scales with the cooperativity $\etafs$.

\section{Interaction between an atomic ensemble and a free-space mode}
\label{Sec:EnsembleFreeSpace}

\subsection{Absorption and dispersion by an ensemble}
\label{Sec:EnsembleAbsorptionFreeSpace}

For an ensemble of $N$ atoms located on the mode axis, the total absorption cross section equals $N$ times the single-atom cross section, Eq. \ref{Eq:CrossSection}, producing Beer's law of exponential attenuation
\begin{equation}\label{Eq:BeersLaw}
    \frac{\Pin - \Pabs}{\Pin}=e^{-\im{2N \beta}}.
\end{equation}
The exponential absorption arises as each layer of atoms is driven by a total field that consists of the incident field on the previous layer and the forward scattered field by that previous layer \cite{FeynmanLecturesI}. If the laser is tuned to atomic resonance,
\begin{equation}\label{Eq:BeersLawResonant}
   \left( \frac{\Pin - \Pabs}{\Pin} \right)_{\omega=\WA}=e^{-2 N \etafs},
\end{equation}
i.e., the resonant ensemble optical depth equals twice the collective cooperativity $N \etafs$.

Similarly, the phase shift induced by the ensemble on the light field is just $N$ times the single-atom phase shift, Eq. \ref{Eq:PhaseShiftFreeSpace}, $\phi_N = N \phi = \re{N \beta}$, and at large detuning $\DA$ from atomic resonance, but within the RWA, $\Gamma \ll \DA \ll \WA$ can be written as
\begin{equation}\label{Eq:PhaseShiftFreeSpaceEtaEnsemble}
    \left( \phi_N \right)_{RWA} = N \etafs \Ld(\DA) \approx - N \etafs \frac{\Gamma}{2\DA}.
\end{equation}

Comparing the absorption and dispersion by a single atom to that by an atomic ensemble, we see that the single-atom cooperativity $\etafs$, Eq. \ref{Eq:FSEta}, for the former is replaced by the collective cooperativity $N \etafs$ for the latter. The fact that the phase shift experienced by the light at fixed light-atom detuning is proportional to the atom number and a geometric parameter can be used for dispersive measurements of atom number or atomic state \cite{Hope04,Hope05,Lodewyck09}, and for measurement-induced spin squeezing in free space \cite{Kuzmich98,Appel09}.

Neither the absorption nor the dispersion depend (with interferometric sensitivity) on the distribution of atoms although both effects rely on a definite phase relationship between the incident field and the forward-scattered field by the atoms. The reason is the cancelation of the phases of the incident and scattered fields in the forward direction: an atom at position $z_1>0$ experiences a drive field whose phase is delayed by $kz_1$ relative to an atom at $z=0$, but the phase of the field emitted forward is advanced by the same amount. Therefore the contributions of all atoms are phase matched in the forward direction, producing maximum interference, independent of the distribution of atoms along the beam. As we shall now see, this is no longer the case when we consider the scattering into a direction other than the direction of the incident beam: The scattered power in any given direction is strongly influenced by the atomic distribution due to interatomic interference.

\subsection{Scattering into a free-space mode by an ensemble: cooperative effects from spatial ordering}
\label{Sec:EnsembleScatteringFreeSpace}

In the geometry of Fig. \ref{Fig:FSScatteringSetup} for scattering from a driving beam into mode $\M$ we assume that the single atom is replaced by $N$ atoms that for simplicity are located at positions ${\bf r}_j$ sufficiently close to the mode axis such that they all couple to $\M$ with the same magnitude. We also assume that the scattered field $\EMN$ in mode $\M$ is small compared to the incident field so that we can take the induced dipoles to be proportional to the incident field $\E$ alone, whose magnitude is assumed to be the same for all atoms (i.e., the sample is optically thin along the incident beam). The phase of the contribution from any atom to the mode amplitude $\EM$ of the radiated field depends on the atom's position, and we can use Eq. \ref{Eq:EMiBetaE} to write
\begin{equation}\label{Eq:ModeAmplitudeCollective}
     \EMN=  i N \FF \beta \E
\end{equation}
in terms of a collective coupling parameter
\begin{equation}\label{Eq:CollectiveCouplingParameter}
     \FF= \frac{1}{N} \sum_{j=1}^N e^{i({\bf k}- \kM) \cdot {\bf r}_j} \equiv \AverDist{e^{i({\bf k}- \kM) \cdot {\bf r}}}.
\end{equation}
Here ${\bf k}$ and $\kM$ are the wavevectors of the incident field and mode $\M$, respectively, and $\{ \}$ denotes the average atomic coupling for the given fixed atomic distribution as defined by Eq. \ref{Eq:CollectiveCouplingParameter}. The power $\PMN=|\EMN|^2/2$ scattered by the ensemble (unidirectionally) into mode $\M$ relative to the power scattered by a single atom into free space $\Pfs=\im{\beta} |\E|^2$ (Eq. \ref{Eq:FSPower4Pi}) is then given by
\begin{equation}\label{Eq:EtaFreeSpaceCollective}
    \frac{\PMN}{\Pfs} = \frac{1}{2} |\FF|^2 N^2 \etafs.
\end{equation}
(Compared to Eq. \ref{Eq:FSMFreeSpacePowerRatio}, here we consider only one direction of $\M$, as in general the factor $\FF$ will be different for the two directions of propagation.) Due to the phase factors in $\FF$ the emission into $\M$ by the ensemble depends on the spatial ordering of the atoms that determines the extent of interference between the fields coherently scattered by different atoms. In particular, $|\FF|^2$ can take on any value between 0 and $1$. The lowest value $|\FF|^2=0$ corresponds to perfect destructive interference between the contributions by different atoms and is, e.g., attained for a perfectly ordered ensemble that contains an integer number $n \geq 2$ of atoms per wavelength. The highest possible value $|\FF|^2=1$ is attained for a periodic lattice of atoms with reciprocal lattice vector ${\bf k} - \kM$, such that the fields emitted by all atoms into $\M$ interfere constructively. This situation corresponds to Bragg scattering \cite{Slama05,Slama05b} and interestingly can arise in a self-organizing manner, due to light forces on the atoms generated by the interference pattern between the incident and the scattered fields \cite{Domokos02,Black03,Zippilli04,Black05,Baumann10}. In this situation the power emitted into $\M$ scales as $N^2$, similar to the situation encountered in superradiance \cite{Dicke54}.

Finally, in the common situation of a gaseous ensemble, corresponding to a random distribution of atoms, $\aver{\FF}=0$ and  $\aver{|\FF|^2}=1/N$, i.e, the phase of the emitted light field is completely random when an ensemble-average over different atomic distributions is performed, and the ensemble-averaged emitted power is proportional to the atom number $N$. The fact that for a random distribution of atoms the emitted power in any given direction is (on average) proportional to the atom number also explains why the usual picture of each atom emitting power independently is valid for gaseous samples, even though in the low-saturation limit all emitted light is coherent, and thus the fields from different atoms interfere. However, we have also seen that the absence of interatomic interference (on average) is just a special, although common, case occurring for disordered ensembles, and that for ordered ensembles both superradiant (emitted power scales as $N^2$) and subradiant (little emitted power) coherent Rayleigh scattering into a given mode is possible.

We have already noted in Section \ref{Sec:EnsembleAbsorptionFreeSpace} that the absorption from a mode does not depend on the atomic distribution, while the emission into a particular mode does. Since the absorbed power must equal the total scattered power by virtue of energy conservation, it follows that cooperative effects in scattering from an (ordered) distribution of atoms correspond merely to a directional redistribution between different free-space modes, and that the total power emitted into free space does not change (see also \cite{Berman10}). In particular, it is not possible to change the scattering cross section per atom by ordering the ensemble. It should be kept in mind, however, that in this argument and in the derivation of the formulas of this section we have assumed that the scattered field in mode $\M$ is much smaller than the driving field ($|\EMN|^2 \ll |\E|^2$), so that we could ignore the backaction of $\EMN$ on the atomic dipoles, and assume that they are driven by the incident field $\E$ alone. When below analyzing the interaction with a cavity mode we will drop this restriction, with interesting consequences.

\section{Interaction between a single atom and a cavity mode} \label{Sec:AtomCavity}

Based on the quantitative understanding of atomic emission into and absorption from a single Gaussian mode in free space we can now analyze the classical interaction between a single atom and a single mode of an optical resonator. In the microwave domain the cavity can partly or completely surround the atom, modifying strongly the total emitted power $\Pfs$ \cite{Kleppner81,Haroche85}. In contrast, the active modes of an optical resonator typically subtend only a very small solid angle. Since we are concerned with optical transitions, we will assume as in the previous section that the solid angle subtended by the cavity mode is small. One might na\"{\i}vely expect that in this case the scattering into free space for a ``driven-atom" setup (Fig. \ref{Fig:CScatteringSetup}) is not affected by the cavity, but as we will see, a cavity supporting a strongly coupled mode can reduce the atomic emission into all free-space modes by acting back on the induced dipole $p=\alpha E$ which depends on the total field $E$ experienced by the atom. This situation arising in a two-level atom driven by two fields is akin to electromagnetically induced transparency (EIT) \cite{Harris89,Harris97} occurring in a three-level atom driven by two fields.

We assume that the atom is at rest and ignore light forces and the photon recoil. A stationary atom that is continuously and weakly driven can be treated as a classical dipole since it simply scatters the incoming narrowband radiation elastically without changing the radiation frequency (coherent or elastic Rayleigh scattering) \cite{Mollow69,API98}. The driven atom inside the optical resonator can then be treated as a monochromatic source of radiation at the frequency $\omega=ck$ of the driving light.

\subsection{Attenuation of a cavity mode: cavity-enhanced absorption} \label{Sec:CavityAbsorption}

\begin{figure}
\centering{\includegraphics[width=5in]{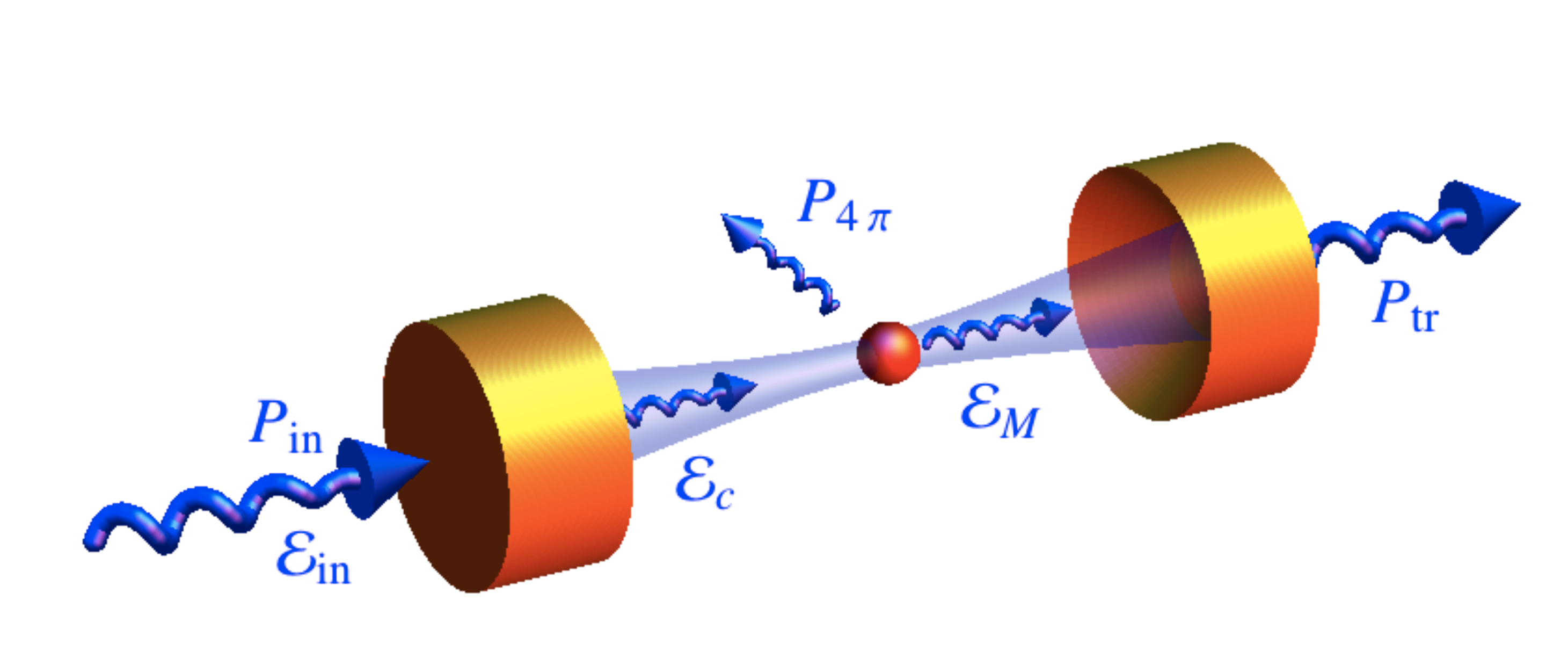}}
\caption{\label{Fig:CTransmissionSetup} Transmission through an optical standing-wave resonator containing an atom. An incident field $\EinMA$ produces a steady-state intracavity field with traveling mode amplitude $\Ec$. The atom at an antinode driven by the field $2 \Ec$ contributes a field $2\EM$ per round trip. The transmitted power is $\Ptr$, the power scattered by the atom into free space is $\Pfs$.}
\end{figure}

We consider a standing-wave resonator of length $L$ with two identical, lossless, partially transmitting mirrors (Fig. \ref{Fig:CTransmissionSetup}) with real amplitude reflection and transmission coefficients $r$ and $iq$, respectively ($r,q$ real, $r^2+q^2=1$), and $q^2 \ll 1$. The resonator supports a TEM$_{00}$ mode with waist size $w$ (mode $\M$), and the atom is located on the mode axis near the waist at an antinode. $\EinMA=\sqrt{\epsn c A}\Ein$ is the mode amplitude incident onto the cavity and $\Ec$ is the mode amplitude of the traveling intracavity field. The mode amplitude leaking into the cavity through the input mirror is $iq \EinMA$, and the atom at the antinode, driven by a field $\E = 2\Ec$, coherently scatters a field $2\EM =4i\beta \Ec$ (see Eq. \ref{Eq:EMiBetaE}) into the resonator that adds to $\Ec$. (The factor of 2 here arises from simultaneous scattering into both cavity directions by the atom at an antinode.) The traveling field $\Ec$ thus experiences reflection at the cavity mirrors, as well as input coupling and atomic source terms, $iq \EinMA$ and $2\EM$, respectively, per roundtrip. The steady-state amplitude $\Ec$ can be determined from the condition that the field after one round trip be unchanged:
\begin{equation}\label{Eq:CavitySelfConsistencyAbsorption}
  \Ec =  r^2 e^{2ikL} \Ec + iq\EinMA+2\EM,
\end{equation}
where $e^{2ikL}$ accounts for the round-trip phase experienced by the circulating light of frequency $\omega=ck$. For not too large detuning $\dc \equiv \omega - \wc \ll  \pi c/L$ from cavity resonance $\wc$, we can approximate $r^2 e^{2ikL} \approx 1- q^2 + 2i q^2 \dc/\kappa $, where $\kappa= q^2c/L$ is the resonator linewidth (decay rate constant of the energy), see e.g., \cite{SiegmanLasers}.

Solving for the cavity field, we find
\begin{equation}\label{Eq:CavityFieldAbsorption}
    \Ec = \frac{i \EinMA}{q} \left[ 1- i\frac{2\dc}{\kappa} - i\frac{4\beta}{q^2} \right]^{-1}.
\end{equation}
The ratio of transmitted power $\Ptr=q^2|\Ec|^2/2$  to incident power $\Pin=|\EinMA|^2/2$ is then
\begin{equation}\label{Eq:CTransmission}
    \frac{\Ptr}{\Pin} = \left[ \left(1+ \frac{\im{4\beta}}{q^2} \right)^2 + \left(\frac{2 \dc}{\kappa} + \frac{\re{4\beta}}{q^2} \right)^2 \right]^{-1},
\end{equation}
Here $\beta=k \alpha /(\pi w^2 \epsn)$, Eq. \ref{Eq:Beta}, containing the atomic polarizability $\alpha$, is evaluated at the frequency $\omega=ck$ of the incident light. The atom can change the transmission through the cavity not only via absorption $\propto \im{\beta} \propto \im{\alpha} $, but also by shifting the cavity resonance via $\re{\beta} \propto \re{\alpha}$, i.e. via the atom's index of refraction that introduces a phase shift of the light (see section \ref{Sec:FSDipsersion}). Both absorptive and dispersive effects can be used for single-atom detection by means of an optical resonator \cite{McKeever04,Hope04,Hope05,Teper06,Puppe07,Trupke07,Poldy08,Heine09,Gehr10,Bochmann10}.

The power $\Pfs$ emitted by the atom into free space is given by Eq. \ref{Eq:FSPower4Pi}, with $\EM=2i\beta \Ec$. The ratio of emitted to incident power $\Pin$ can be written as
\begin{equation}\label{Eq:CFreeSpaceEmissionCavityDrive}
    \frac{\Pfs}{\Pin} =\frac{\im{8\beta}}{q^2} \left[ \left(1+ \frac{\im{4\beta}}{q^2} \right)^2 + \left(\frac{2 \dc}{\kappa} + \frac{\re{4\beta}}{q^2} \right)^2 \right]^{-1},
\end{equation}

In the RWA, the coupling factor $\beta$ takes the simple form of Eq. \ref{Eq:BetaRWA}, and we can write
\begin{equation}\label{Eq:BetaCavityRWA}
    \left( \frac{4\beta}{q^2} \right)_{RWA} = \etac \left( \Ld(\DA) + i \La(\DA) \right),
\end{equation}
where we have defined a cavity cooperativity parameter (also called the Purcell factor \cite{Purcell46,Motsch10})
\begin{equation}\label{Eq:EtaCavity}
     \etac = \frac{4 \etafs}{q^2} = \frac{24}{q^2 k^2 w^2} = \frac{24 \F /\pi}{k^2 w^2}.
\end{equation}
Here $\F = \pi c/(L \kappa)=\pi/q^2$ is the cavity finesse, and $\La(\DA)=\Gamma^2/(\Gamma^2+4 \Delta^2)$ and $\Ld(\DA)=-2 \DA \Gamma /(\Gamma^2+4 \Delta^2)$ are the Lorentzian absorptive and dispersive lineshapes, respectively. The cavity cooperativity can be understood as the free-space cooperativity $\etafs$ augmented by the average number of photon round trips $\F/ \pi$ inside the cavity, with an additional factor of four accounting for the four times larger intensity at an antinode of a standing wave compared to a traveling mode. (Note also that the above defined cooperativity parameter $\eta$ is twice as large as the cooperativity parameter $C_1$ most widely used in cavity QED, see, e.g., \cite{Horak03}.)

Eq. \ref{Eq:BetaCavityRWA} can be substituted into Eqs. \ref{Eq:CTransmission}, \ref{Eq:CFreeSpaceEmissionCavityDrive} to write explicit expressions in the RWA for the resonator transmission and free-space emission as a function of cavity cooperativity $\etac$, detuning between the incident light and the cavity resonance $\delta=\omega - \wc$, and detuning between the incident light and the atomic resonance $\Delta = \omega-\WA$:
\begin{equation}\label{Eq:CTransmissionRWA}
   \left( \frac{\Ptr}{\Pin} \right)_{RWA}= \frac{1}{\left[1+ \etac \La(\DA) \right]^2 + \left[\frac{2 \dc}{\kappa} + \etac \Ld(\DA) \right]^2 }
\end{equation}
and
\begin{equation}\label{Eq:CFreeSpaceEmissionCavityDriveRWA}
    \left( \frac{\Pfs}{\Pin} \right)_{RWA} = \frac{2 \etac \La(\DA)}{\left[1+  \etac \La(\DA) \right]^2 + \left[\frac{2 \dc}{\kappa} + \etac \Ld(\DA) \right]^2}.
\end{equation}
Similar expressions were already derived by \cite{Zhu90} with a classical formalism as used here, and they agree with the quantum mechanical formulas in the low-saturation limit. Atomic absorption, spectrally characterized by the absorptive Lorentzian $\La(\DA)$ and scaled by the cavity cooperativity parameter $\etac$, reduces the intracavity power and the transmission, while Lorentzian atomic dispersion $\etac \Ld(\DA)$ shifts the cavity resonance. In the expression for the free-space emission, Eq. \ref{Eq:CFreeSpaceEmissionCavityDriveRWA}, the absorptive Lorentzian appears also in the numerator since for a given intracavity power the atomic free-space emission scales in the same way as the absorption.

\begin{figure}
\centering{\includegraphics[width=4in]{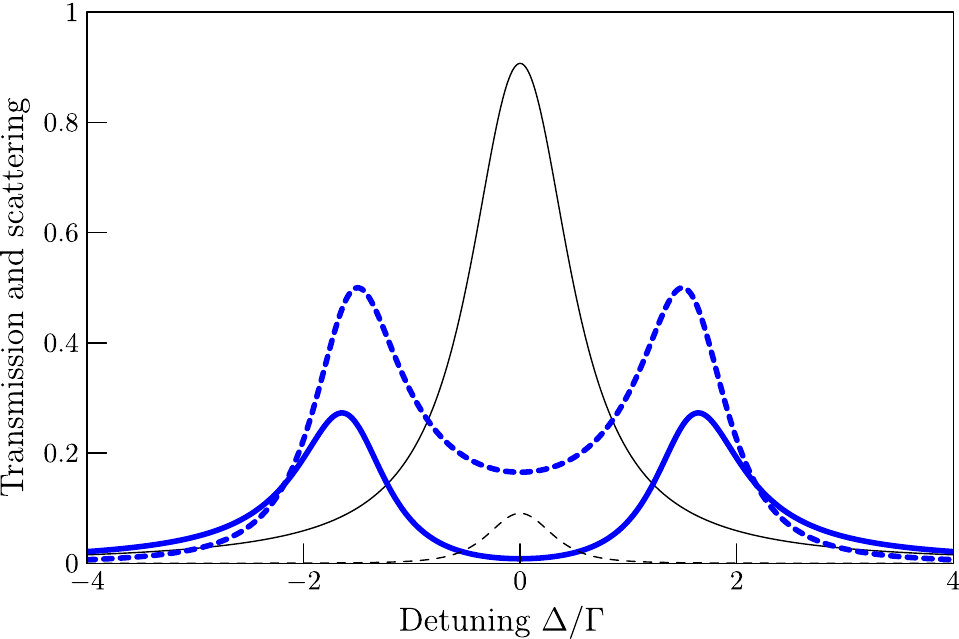}}
\caption{\label{Fig:CTransmission} Transmission through the cavity (solid) and free-space scattering (dashed) for a resonant atom-cavity system ($\wc=\WA$) vs. detuning $\DA=\dc$ in units of $\kappa=\Gamma$ for a weakly coupled system ($\etac=0.05$, thin black lines) and for a strongly coupled system ($\etac=10$, thick blue lines). Both transmission and scattering are normalized to the power incident on the cavity. The strongly coupled system exhibits vacuum Rabi splitting, i.e. the normal-mode splitting exceeds the normal-mode widths.}
\end{figure}

\begin{figure}
\centering{\includegraphics[width=4in]{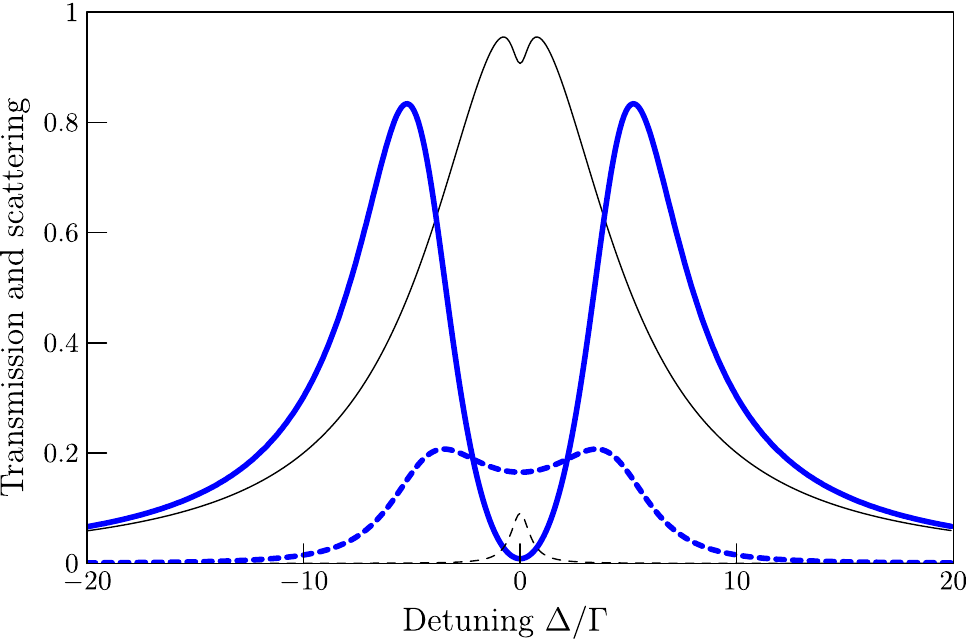}}
\caption{\label{Fig:CTransmissionA} Transmission through the cavity (solid) and free-space scattering (dashed) for a resonant atom-cavity system ($\wc=\WA$) vs. detuning $\DA=\dc$ in units of $\Gamma$ for $\kappa=10\Gamma$ for a weakly coupled system ($\etac=0.05$, thin black lines) and for a strongly coupled system ($\etac=10$, thick blue lines). Both transmission and scattering are normalized to the power incident on the cavity. In this situation there is no standard Rabi splitting as the cavity width is larger than the normal-mode splitting, but the transmission drops sharply near $\DA=0$, akin to the situation in EIT.}
\end{figure}

The transmission and scattering into free space are plotted as a function of incident frequency $\omega$ for fixed cavity frequency in a few representative cases in Figs. \ref{Fig:CTransmission}, \ref{Fig:CTransmissionA}. For $\etac < 1$ (weak-coupling limit) the atomic absorption broadens the linewidth and reduces the transmission, while the atomic dispersion induces a cavity shift. In the weak-coupling limit the two eigenmodes of the system, one atom-like, the other cavity-like, maintain their character, each with a little admixture of the other mode. In the opposite strong-coupling limit $\eta>1$ the two modes are strongly mixed when the cavity resonance coincides with the atomic resonance. Both cavity transmission and atomic emission into free space show a normal-mode splitting, given by $2g=\sqrt{\eta \Gamma \kappa}$, that in the quantum description for $2g>(\Gamma,\kappa)$ is interpreted as the vacuum (or single-photon) Rabi splitting of cavity QED \cite{Kimble98,API98}.

In the classical picture the single-photon Rabi splitting or normal-mode splitting for a resonant atom-cavity system ($\wc=\WA$, i.e. $\dc=\DA$) and similar cavity and atomic linewidths ($\kappa \sim \Gamma$) can be understood as follows (Fig. \ref{Fig:CTransmission}): On resonance for $\etac>1$ the atomic absorption spoils the cavity finesse, and the intracavity and transmitted power are low. As the laser is detuned away from resonance, the atomic absorption is reduced and the transmission increases until the cavity loss due to atomic emission no longer limits the remaining constructive interference arising from multiple round trips of the light in the detuned cavity. (The round trip phase also includes the atomic contribution that has the opposite sign as the cavity contribution and tends to decrease the total roundtrip phase, and increase the intracavity power.) Further detuning $|\dc|$ then again decreases the intracavity power as the increasing round-trip phase shift decreases the constructive interference inside the cavity. The combination of atomic absorption and dispersion results in two transmission peaks that are symmetric about $\delta=0$.

If the atomic linewidth is much narrower that the cavity linewidth ($\Gamma \ll \kappa$) then the atomic absorption affects the cavity transmission only in a narrow region near atomic resonance (Fig. \ref{Fig:CTransmissionA}). The transmission is substantially reduced for $\eta>1$, but if the cooperativity parameter is not too large ($\etac < \kappa/\Gamma$) the normal-mode splitting is less than the cavity linewidth, and there is no standard Rabi splitting. Rather, there is a dip in the transmission and in the free-space scattering.

The ratio of atomic free-space scattering to cavity transmission is given by the simple expression
\begin{equation}\label{Eq:EmissionTransmissionRatio}
     \left( \frac{\Pfs}{\Ptr} \right)_{RWA}= 2\etac \La(\DA),
\end{equation}
and independent of the atom-cavity detuning $\dc-\DA$. For a resonant system ($\dc=\DA=0$) the transmission and free-space scattering are given by
\begin{equation}\label{Eq:CResonantTransmission}
     \left( \frac{\Ptr}{\Pin} \right)_{\DA=\dc=0}=\frac{1}{(1+\etac)^2},
\end{equation}
and
\begin{equation}\label{Eq:CResonantTransmissionScattering}
     \left( \frac{\Pfs}{\Pin} \right)_{\DA=\dc=0}=\frac{2\etac}{(1+\etac)^2}.
\end{equation}
Comparison of Eq. \ref{Eq:CResonantTransmissionScattering} to the corresponding free-space equation \ref{Eq:FSAbsorption} shows that in the weak-coupling limit $\etac<1$ the quantity $2\etac=8 \F \etafs/\pi$ can be interpreted as the cavity-aided optical depth. In the strong-coupling limit $\etac \gg 1$ both the transmission and the free-space scattering decrease with coupling strength $\etac$, but the transmission decreases faster than the free-space scattering. This is closely related to EIT \cite{Harris89,Harris97} where the population of the state or mode driven by the probe field (here the resonator, in EIT the atomic excited state) is more suppressed than that of the indirectly driven state or mode (here the free-space modes, in EIT the outer atomic ground state).

Eq. \ref{Eq:CResonantTransmission} also shows that in a cavity the transmitted power decreases only quadratically, rather than exponentially, with optical depth $2 \etac>1$. The reason is that the enhanced absorption resulting in $\etac = (4\F/\pi) \etafs$ is due to multiple round trips inside the cavity: as the atomic absorption per round trip increases, the cavity finesse $\F$ and the number of round trips $\F/\pi$ decrease, which acts to convert the exponential absorption into a polynomial one. (The single-pass optical depth is $2\etafs=12/(k^2 w^2)<1$.)

\subsection{Frequency shift of a cavity mode: dispersion}

In the limit of sufficiently large detuning from atomic resonance, such that the cavity finesse is not spoiled by atomic absorption ($\etac \La(\DA) < 1$), the dominant effect of the atom on the resonator is a shift of the cavity resonance frequency by atomic dispersion, since the real part of the atomic polarizability falls off more slowly with detuning than the imaginary part. From Eq. \ref{Eq:CTransmissionRWA} it follows that the atom-induced cavity resonance shift $\dac$, in units of the cavity linewidth $\kappa$, in the RWA is given by
\begin{equation}\label{Eq:CavityLineShift}
   \left( \frac{\dac}{\kappa} \right)_{RWA} = -\frac{\etac}{2} \Ld(\DA) \approx \etac \frac{\Gamma}{4 \DA},
\end{equation}
which is proportional to the cavity cooperativity parameter $\etac$. The atom-induced cavity shift can be used for atom detection or atomic-state detection \cite{McKeever04,Hope04,Hope05,Teper06,Puppe07,Trupke07,Poldy08,Heine09,Terraciano09,Gehr10,Bochmann10}, or, in the case of an atomic ensemble, for generating cavity-mediated infinite-range atomic-state-dependent interactions between atoms enabling spin squeezing \cite{Schleier-Smith10a,Leroux10,Leroux10a}.

\subsection{Scattering into a cavity mode: cavity-enhanced emission} \label{Sec:CavityScattering}

\begin{figure}
\centering{\includegraphics[width=5in]{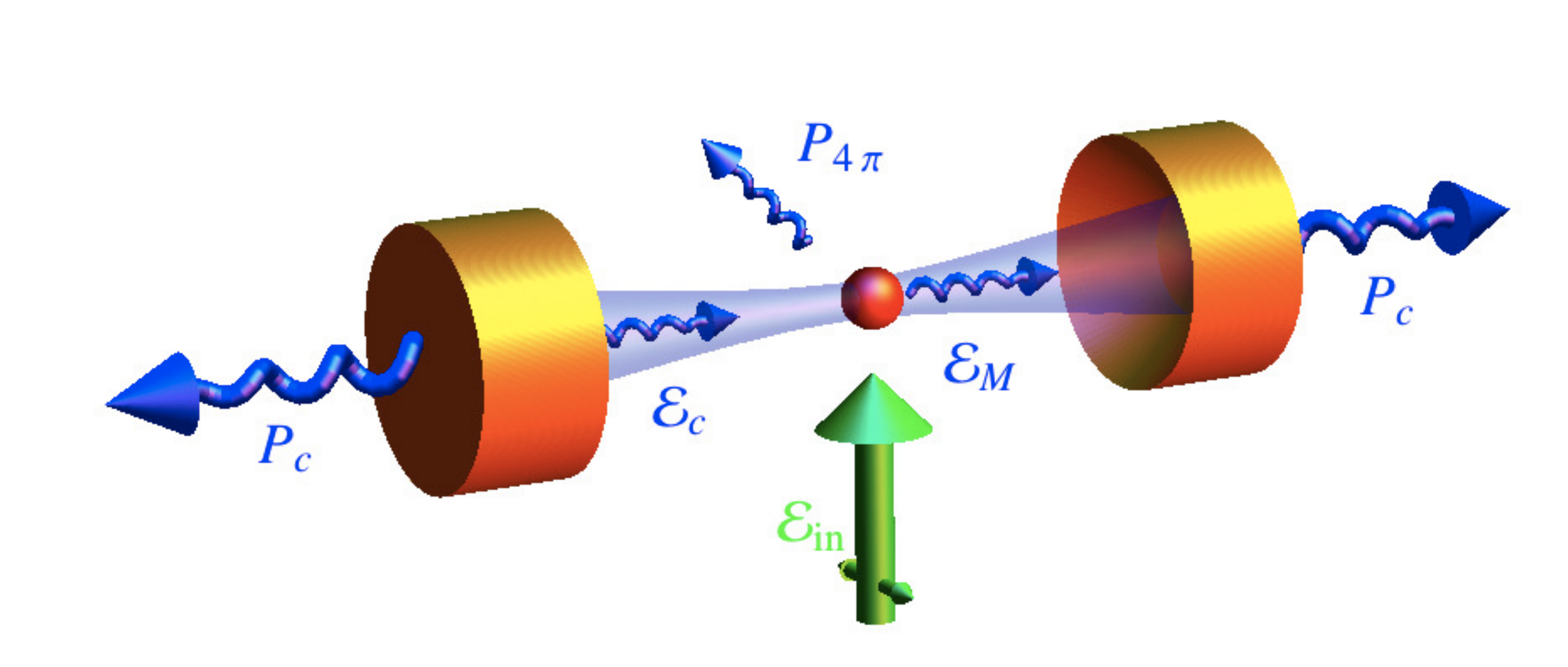}}
\caption{\label{Fig:CScatteringSetup} An atom driven by an incident field $\EinMA$ scattering monochromatic radiation into an optical standing-wave resonator. The traveling mode amplitude is $\Ec$, the atom at an antinode adds a mode amplitude $2 \EM$ per round trip. The power leaving the cavity in both directions is $\Pc$, the power scattered by the atom into free space is $\Pfs$.}
\end{figure}

We now consider the scattering of radiation by an atom into a resonator of the same geometry and parameters as in Section \ref{Sec:CavityAbsorption}. The atomic dipole is driven by a mode amplitude $\EinMA$ of frequency $\omega=ck$ from the side, and emits monochromatic radiation of the same frequency into the resonator (Fig. \ref{Fig:CScatteringSetup}). In particular, the atom at an antinode contributes a mode amplitude $2\EM$ per round trip to the mode amplitude $\Ec$ of the circulating field inside the cavity. In steady state, $\Ec$ can be determined from the condition that the field after one round trip, experiencing reflection at the mirrors as well as the atomic source term, be unchanged \cite{SiegmanLasers}:
\begin{equation}\label{Eq:CavitySelfConsistencyScattering}
  \Ec =  r^2 e^{2ikL} \Ec + 2\EM ,
\end{equation}
which, under the same conditions as in Section \ref{Sec:CavityAbsorption} (not too high mirror transmission $q^2 \ll 1$ and not too large detuning from cavity resonance $\dc \ll c/(2L)$), has a solution of the form
\begin{equation}\label{Eq:CavityFieldSimple}
    \Ec =  \frac{2 \EM}{q^2} \frac{1}{1-2i \dc/ \kappa}.
\end{equation}

The power emitted by the atom into the cavity is determined by the field leaking out through both cavity mirrors, $\Pc= q^2|\Ec|^2$. The power emitted into free space is $\Pfs=|\EM|^2/\etafs$ (Eq. \ref{Eq:FSPower4Pi}) and using Eq. \ref{Eq:EtaCavity} the ratio of cavity-to-free-space emission can be simply written as
\begin{equation}\label{Eq:CavityEmissionRatioSimple}
    \frac{\Pc}{\Pfs} = \etac \frac{\kappa^2}{\kappa^2+ 4\dc^2}.
\end{equation}
Compared to the emission into the same free-space mode $\etafs$, as given by Eq. \ref{Eq:FSEta}, the resonant cavity ($\dc=0$) enhances the emission by a factor $4/q^2=4\F/\pi$. This factor arises from the constructive interference between the images of the atomic dipole formed by the cavity mirrors, or equivalently, from the constructive interference of the atomic emission on successive round trips of the light during the lifetime of the cavity. This frequency-dependent enhancement of coherent scattering that persists even at large detuning from atomic resonance, as observed by \cite{Motsch10}, is the principle behind cavity cooling \cite{Mossberg91,Horak97,Vuletic00,Vuletic01,Maunz04,Nussmann05,Zippilli05a,Morigi07,Lev08,Leibrandt09}.

Note the formal similarity between the result for cavity emission by the driven atom (Eq. \ref{Eq:CavityEmissionRatioSimple}) and free-space emission when the cavity is driven (Eq. \ref{Eq:EmissionTransmissionRatio}): Apart from the factor of 2 difference between absorption and scattering (compare also Eqs. \ref{Eq:FSMFreeSpacePowerRatio}, \ref{Eq:FSAbsorptionRWA} for scattering and absorption in free space) the roles of the cavity field and the atomic emission are interchanged in the two cases, and so are the corresponding Lorentzian factors.

While the ratio between cavity ($\Pc$) and free-space ($\Pfs$) emission is independent of atomic parameters and detuning relative to atomic resonance, the individual terms $\Pc$ and $\Pfs$ depend on the atomic polarizability at the frequency of the driving light. To obtain a solution that remains valid in the limit of strong light-atom coupling (large cooperativity $\etac>1$), we need to take self-consistently into account that the atomic dipole ($\propto \EM$) is driven not only by the external field $\EinMA$ but also by the field $\Ec$ of the same frequency circulating inside the cavity. An atom at an antinode experiences a total field $\E = \EinMA + 2\Ec$, and we write Eq. \ref{Eq:EMiBetaE} as
\begin{equation}\label{Eq:ModeAmplitudeCavityDrive}
    \EM = i \beta \left( \EinMA + 2 \Ec \right).
\end{equation}
Substituting $\EM$ into the steady-state condition for the cavity field $\Ec$, Eq. \ref{Eq:CavitySelfConsistencyScattering}, and solving for $\Ec$, we find
\begin{equation}\label{Eq:CavityFieldwithBackactionScattering}
    \Ec =  \frac{2 i \beta \EinMA}{q^2} \frac{1}{1- i\frac{2\dc}{\kappa} - i\frac{4\beta}{q^2}}.
\end{equation}
We can now also find the atomic source term $\EM$ (driven by both incident and cavity fields) by substituting $\Ec$ into Eq. \ref{Eq:ModeAmplitudeCavityDrive} for the atomic dipole,
\begin{equation}\label{Eq:ModeAmplitudeBackaction}
    \EM =  i \beta \EinMA \frac{1-i \frac{2 \dc}{\kappa}}{1- i\frac{2\dc}{\kappa} - i\frac{4\beta}{q^2}}.
\end{equation}
The bidirectional cavity emission rate $\Pc=q^2 |\Ec|^2$, relative to the power emitted into free space in the absence of the cavity $\Pzerofs=|\beta \EinMA|^2/\etafs$, Eq. \ref{Eq:FSPower4Pi}, is then
\begin{equation}\label{Eq:CavityEmissionBackaction}
    \frac{\Pc}{\Pzerofs} = \frac{\etac}{\left(1+ \frac{\im{4\beta}}{q^2} \right)^2 + \left(\frac{2 \dc}{\kappa} +\frac{\re{4 \beta}}{q^2} \right)^2}.
\end{equation}

The emission into free space $\Pfs=|\EM|^2/\etafs$ is similarly modified by the presence of the cavity from its value $\Pzerofs$ in the absence of the cavity:
\begin{equation}\label{Eq:FreeSpaceEmissionBackaction}
    \frac{\Pfs}{\Pzerofs} = \frac{ 1+\left( \frac{2\dc}{\kappa}\right)^2}{\left(1+ \frac{4\im{\beta}}{q^2} \right)^2 + \left(\frac{2 \dc}{\kappa} +\frac{4 \re{\beta}}{q^2} \right)^2}.
\end{equation}
It is highly interesting to see that power emitted into free space can be enhanced or reduced by a cavity that subtends only a tiny solid angle, as has been first noted by \cite{Alsing92} using a quantum mechanical description. The modification of free-space emission is not a saturation effect of the atom, as we have explicitly constructed a classical model that does not include atomic saturation. Rather, it is the backaction of the cavity field driving the atomic dipole in antiphase with the incident field, which reduces the magnitude of the dipole, and thus the amount of emission into free space.

On atomic and cavity resonance ($\dc=\DA=0$) the emission into the cavity and into free space are given by the simple expressions
\begin{equation}\label{Eq:ResonantCavityEmission}
   \left( \frac{\Pc}{\Pzerofs} \right)_{\dc=\DA=0} =  \frac{\etac}{\left(1+\etac \right)^2}
\end{equation}
and
\begin{equation}\label{Eq:ResonantCavityFreeSpaceEmission}
   \left( \frac{\Pfs}{\Pzerofs} \right)_{\dc=\DA=0} =  \frac{1}{\left(1+\etac \right)^2},
\end{equation}
respectively. Note again the complementarity between these formulas and Eqs. \ref{Eq:CResonantTransmission}, \ref{Eq:CResonantTransmissionScattering} for the driven cavity. Although we are considering here only a two-level atom, these formulas are closely related to electromagnetically induced transparency in a three-level system \cite{Harris89,Harris97} as both the incident light and the light inside the cavity couple to the atomic excited state \cite{Field93,Rice96}. The intracavity field builds up $\pi$ out of phase with the driving field at the location of the atom, and acts to reduce the emission by the atom, both into the cavity and into free space \cite{Heinzen87,Alsing92,Zippilli04}.  In the limit of strong coupling $\etac \gg 1$, the intracavity electric field experienced by the atom, $2\Ec \approx -\EinMA$ is independent of the atomic or cavity properties, and builds up to be (almost) equal in value to the driving field at the position of the atom. This reduces the atomic emission into free space by $(1+\etac)^2$, and the dominant emission process is into the cavity. A cavity with perfectly reflecting mirrors ($\etac \raw \infty $) would cancel all resonant free-space emission, even when it subtends only a small solid angle $\Delta \Omega \ll 1$ \cite{Alsing92}.

\begin{figure}
\centering{\includegraphics[width=4in]{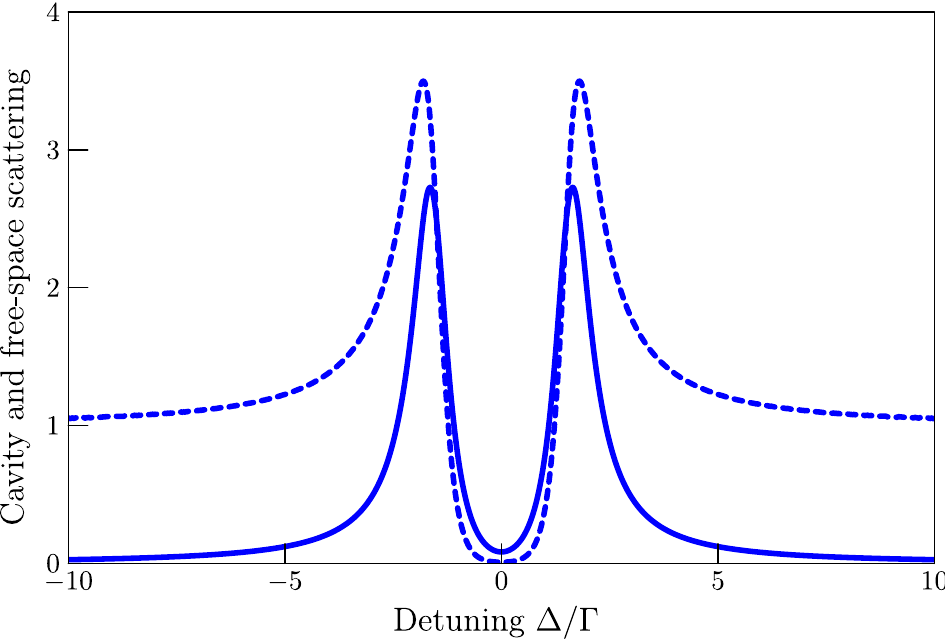}}
\caption{\label{Fig:CScatteringResults} Scattering rate into the cavity $\Pc/\Pzerofs$ (solid line) and into free space $\Pfs/\Pzerofs$ (dashed line) for a cavity resonant with the atomic transition ($\wc = \WA$) vs. probe laser detuning $\dc=\DA$ in units of $\Gamma=\kappa$. The displayed curves are for cooperativity parameter $\etac=10$. Note the suppression of free-space scattering (and cavity scattering) on resonance, and the enhancement of free-space and cavity scattering off resonance. The strong modification of free-space scattering by a cavity subtending only a very small solid angle arises from the interference between the cavity field and the incident field at the atom's position.}
\end{figure}

In the RWA we can substitute Eq. \ref{Eq:BetaCavityRWA} to write explicit expressions for the cavity and free-space scattering as a function of laser frequency:
\begin{equation}\label{Eq:CavityEmissionBackactionRWA}
    \frac{\Pc}{\Pzerofs} = \frac{\etac}{\left[1+ \etac \La(\DA) \right]^2 + \left[\frac{2 \dc}{\kappa} +\etac \Ld(\DA) \right]^2}
\end{equation}
and
\begin{equation}\label{Eq:FreeSpaceEmissionBackactionRWA}
    \frac{\Pfs}{\Pzerofs} = \frac{ 1+\left( \frac{2\dc}{\kappa}\right)^2}{\left[1+ \etac \La(\DA) \right]^2 + \left[\frac{2 \dc}{\kappa} +\etac \Ld(\DA) \right]^2}.
\end{equation}

Both quantities are plotted in Fig. \ref{Fig:CScatteringResults} vs. detuning of the incident laser when the cavity resonance is chosen to coincide with the atomic resonance (i.e., $\wc = \WA$, $\DA=\dc$). For strong atom-cavity coupling, $\etac \gg 1$, both cavity and free-space emission display two maxima split by $2g=\sqrt{\etac \Gamma \kappa}$, i.e. the system shows the normal-mode splitting usually associated with the vacuum Rabi splitting of cavity QED \cite{API98}. We see that this feature appears in linear dispersion theory also when the coupled atom-cavity system is not probed via transmission through the cavity (section \ref{Sec:CavityAbsorption}), but via excitation of the atom.

\begin{figure}
\centering{\includegraphics[width=4in]{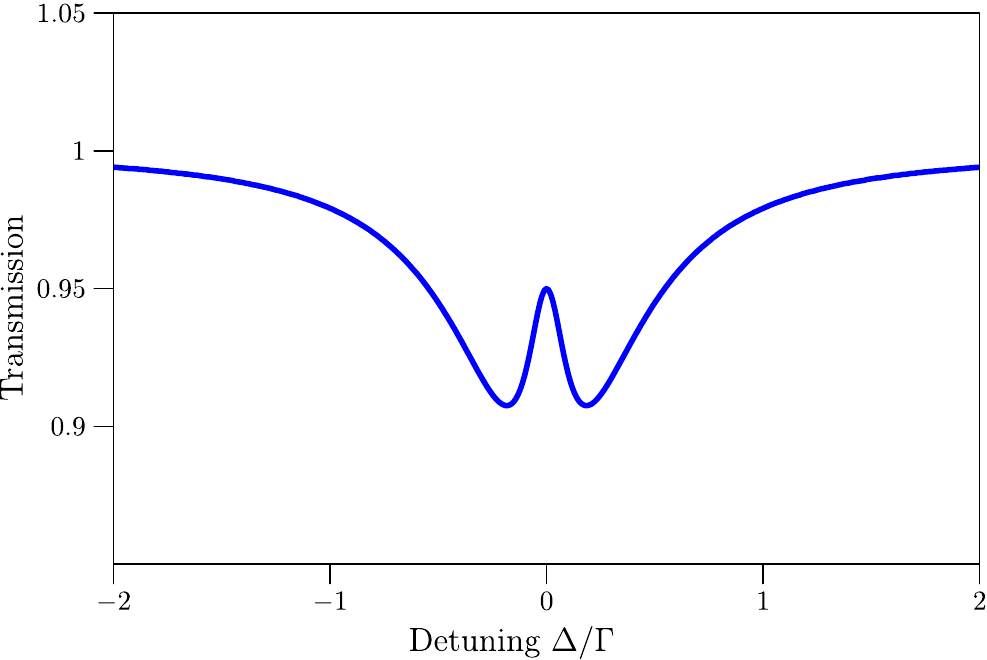}}
\caption{\label{Fig:CavityEIT} Transmission of the sidebeam in the ''scattering scenario" of Fig. \ref{Fig:CScatteringSetup} in the presence of the cavity for a resonant atom-cavity system ($\WA=\wc$) as a function of detuning $\DA$ of the sidebeam in units of $\Gamma$. The cavity linewidth is chosen much narrower than the atomic linewidth, $\kappa=\Gamma/10$, and the absorption of the sidebeam in the absence of the cavity is chosen to be 10\%. The origin of the resonant transmission peak is the same as in EIT, with the strongly coupled cavity ($\etac=1$) replacing the coupling laser in standard EIT.}
\end{figure}

It is interesting to consider the transmission of the beam from the side, $T=1-(\Pfs+\Pc)/\Pin$, which can be calculated from Eqs. \ref{Eq:CavityEmissionBackactionRWA}, \ref{Eq:FreeSpaceEmissionBackactionRWA} and $\Pin=|\EinMA|^2/2$. The sidebeam transmission, displayed in Fig. \ref{Fig:CavityEIT}, for $\kappa < \Gamma$ and $\etac \geq 1$ shows a cavity-induced transmission window within the atomic absorption line. The physical mechanism is the same as in EIT \cite{Harris89,Harris97}, where the strongly coupled cavity mode replaces the usual classical coupling beam \cite{Field93,Rice96}.

In summary, we find that the cooperativity parameter $\etac$ governs the strength the atom-cavity interaction: the fractional scattering into a resonant cavity, the reduction in cavity transmission, and the dispersive shift of the cavity resonance frequency are all determined by the dimensionless factor $\etac$. This factor is the product of the resonant single-pass absorption of the light, as given by the ratio of atomic cross section and beam area, and the average number of photon round trips in the optical resonator, as determined by the cavity finesse $\F$. Since the latter depends only on mirror properties, we find that all resonators with the same mirror reflectivity and the same waist size produce the same strength of atom-light interaction $\etac$, independent of the length of the cavity. In other words, the atom-light interaction, at least in aspects that can be described classically, depends on mode \emph{area}, rather than mode \emph{volume}. Any volume-dependent effects enter through the ratio $\kappa/\Gamma$ of cavity to atomic linewidth, but the classical strong-coupling condition $\eta>1$ is determined by mode area and cavity finesse alone.

\section{Interaction between an atomic ensemble and a cavity mode} \label{Sec:EnsembleCavity}

\subsection{Absorption and dispersion by an ensemble in a cavity mode} \label{Sec:EnsembleCavityAbsorption}

As in the free space case, Section \ref{Sec:EnsembleScatteringFreeSpace}, we consider $N$ atoms located at positions ${\bf r}_j$ sufficiently close to the cavity axis such that the radial variation of the coupling may be ignored (see Fig. \ref{Fig:CTransmissionSetup}). The cavity is driven by an incident field. An atom at an antinode experiences a cavity mode amplitude $2\Ec$ (see Section \ref{Sec:CavityAbsorption}), and hence the atomic source term is
\begin{equation}\label{Eq:EnsembleMode}
   2\EM = 4 i \beta N H \Ec
\end{equation}
with the collective coupling parameter
\begin{equation}\label{Eq:CEnsembleCouplingCavity}
   H= \frac{1}{N} \sum_{j=1}^N \cos^2 k z_j \equiv \AverDist{\cos^2 kz}.
\end{equation}
With the cavity oriented along the z-axis, the cavity field at position $z_j$ driving the dipole is proportional to $\cos kz_j$, and so is the field emitted by the atom into the cavity mode for a given dipole, hence the $\cos^2 k z_j$ dependence. As in Section \ref{Sec:EnsembleScatteringFreeSpace}, the curly brackets denote the average for a given and fixed atom distribution. Solving the steady-state condition for the cavity field, Eq. \ref{Eq:CavitySelfConsistencyAbsorption}, with this atomic source term $2\EM$ from Eq. \ref{Eq:EnsembleMode}, we find for the ratio of transmitted to incident power
\begin{equation}\label{Eq:CavityTransmissionEnsemble}
    \frac{\Ptr}{\Pin} = \left[ \left(1+ \frac{\im{4 N H\beta}}{q^2} \right)^2 + \left(\frac{2 \dc}{\kappa} + \frac{\re{4 N H\beta}}{q^2} \right)^2 \right]^{-1}.
\end{equation}
Since the summands in $H$ are all positive quantities, the result depends only weakly on the ordering of the atoms. A perfectly ordered ensemble with all atoms at antinodes has $H=1$, while a random distribution of atoms along the cavity standing wave has $\aver{H}=\frac{1}{2}$ when averaged over different atomic spatial distributions.

For the total scattering into all free-space modes, there is no interference between different atoms (see section \ref{Sec:EnsembleScatteringFreeSpace}), and the total emitted power is obtained by adding the emitted power of all atoms, Eq. \ref{Eq:FSPower4Pi}. This yields $\PfsN = \im{4 \beta} |\Ec|^2 N H$, and
\begin{equation}\label{Eq:FreeSpaceEmissionCavityDriveEnsemble}
    \frac{\PfsN}{\Pin} =N H \frac{\im{8 \beta}}{q^2} \left[ \left(1+ \frac{\im{4 N H \beta}}{q^2} \right)^2 + \left(\frac{2 \dc}{\kappa} + \frac{\re{4 N H \beta}}{q^2} \right)^2 \right]^{-1}.
\end{equation}

In the RWA we can write for the transmission and free-space scattering
\begin{equation}\label{Eq:CEnsembleTransmissionRWA}
   \left( \frac{\Ptr}{\Pin} \right)_{RWA}= \frac{1}{\left[1+ H N \etac \La(\DA) \right]^2 + \left[\frac{2 \dc}{\kappa} + H N \etac \Ld(\DA) \right]^2 }
\end{equation}
and
\begin{equation}\label{Eq:CEnsembleFreeSpaceEmissionCavityDriveRWA}
    \left( \frac{\Pfs}{\Pin} \right)_{RWA} = \frac{2 H N \etac \La(\DA)}{\left[1+  H N \etac \La(\DA) \right]^2 + \left[\frac{2 \dc}{\kappa} + H N \etac \Ld(\DA) \right]^2}.
\end{equation}

Comparison of these equations to Eqs. \ref{Eq:CTransmissionRWA}, \ref{Eq:CFreeSpaceEmissionCavityDriveRWA} shows that for the ensemble the single-atom cooperativity $\etac$ is replaced by the collective cooperativity $N \etac$, with a proportionality factor between 0 and 1, given by $H=\AverDist{\cos^2 kz}$, that depends on the atomic distribution relative to the cavity standing wave. Similarly, the cavity shift at large detuning from atomic resonance in the RWA, $\WA \gg \DA \gg \Gamma$ is given by
\begin{equation}\label{Eq:CResonatorShiftEnsemble}
   \left( \frac{\dac}{\kappa} \right)_{RWA} = -\frac{1}{2} H N \etac \Ld(\DA) \approx H N \etac \frac{\Gamma}{4 \DA}.
\end{equation}
Since $H=\AverDist{\cos^2 kz}$ depends only weakly on the atomic distribution as it varies from a disordered ($\aver{H}=\frac{1}{2}$) to a superradiant ($H=1$) situation, one does not expect the atomic trajectories to influence each other severely \cite{Domokos01}. The situation is very different if the system is excited from the side, i.e. if the cavity mode is excited via the atomic scattering, as discussed in the next section.

\subsection{Scattering by an ensemble into a cavity mode} \label{Sec:EnsembleCavityScattering}

We consider an ensemble of $N$ atoms at positions ${\bf r}_j$ in a cavity oriented along $z$, as in the previous Section \ref{Sec:EnsembleCavityAbsorption}, but now being driven with a beam from the side traveling along $x$, as in Fig. \ref{Fig:CScatteringSetup}. The ensemble is assumed to be optically thin for the incident field so that all atoms experience the same incident-field magnitude. As each atom is driven both by the incident field ($\Ein$) and the cavity mode ($2\Ec$ at an antinode), the atomic source term is
\begin{equation}\label{Eq:CEnsembleModeSidebeam}
   2\EM = 2 i \beta N \left( G \EinMA + 2H \Ec \right)
\end{equation}
with the collective coupling parameter $H=\AverDist{\cos^2 kz}$ along the cavity given by Eq. \ref{Eq:CEnsembleCouplingCavity}, and the collective coupling parameter for the incident beam being
\begin{equation}\label{Eq:CEnsembleCouplingIncident}
   \GG= \frac{1}{N} \sum_{j=1}^N e^{i k x_j}\cos k z_j \equiv \AverDist{e^{i k x}\cos k z}.
\end{equation}
Using the same procedure as in Section \ref{Sec:CavityScattering}, i.e. inserting the expression for $\EM$ into the steady-state condition for the cavity field, Eq. \ref{Eq:CavitySelfConsistencyScattering}, and solving for $\Ec$, we have now
\begin{equation}\label{Eq:CavityFieldwithBackactionScatteringEnsemble}
    \Ec =  \frac{2 i \beta N \GG \EinMA}{q^2} \frac{1}{1- i\frac{2\dc}{\kappa} - i\frac{4 N H \beta}{q^2}}.
\end{equation}
This yields for the power scattered into the cavity relative to the power $\Pzerofs$ scattered by a single atom into free space in the absence of the cavity
\begin{equation}\label{Eq:CEnsembleScattering}
    \frac{\PcN}{\Pzerofs} =\frac{|\GG|^2 N^2 \etac}{\left(1+ \frac{\im{4 N H \beta}}{q^2} \right)^2 + \left(\frac{2 \dc}{\kappa} + \frac{\re{4 N H \beta}}{q^2} \right)^2}.
\end{equation}
In the RWA we can use Eq. \ref{Eq:BetaRWA} to write for the scattering into the cavity
\begin{equation}\label{Eq:CavityEmissionBackactionEnsembleRWA}
    \frac{\PcN}{\Pzerofs} = \frac{|\GG|^2 N^2 \etac }{\left[1+ H N \etac \La(\DA) \right]^2 + \left[\frac{2 \dc}{\kappa} + H N \etac \Ld(\DA) \right]^2},
\end{equation}
The atomic distribution along the cavity axis as quantified by $H=\AverDist{\cos^2kz}$ determines the absorption and dispersion of the resonator, while the distribution with respect to both the incident beam and the cavity as quantified by $\GG=\AverDist{e^{i k x}\cos k z}$ determines the scattering into the resonator. If the atomic detuning $\DA$ is large enough that the absorption can be ignored ($H N \etac \La(\DA) < 1$), then the scattering into the cavity can have super- or subradiant features similar to those discussed for the free-space case in Section \ref{Sec:EnsembleScatteringFreeSpace}. In particular, for an average over randomly ordered ensembles we have $\aver{|\GG|^2}=\frac{1}{2N}$, i.e. the scattering into the cavity is proportional to the atom number, while for a perfectly ordered ensemble $\GG=1$, i.e. the emission into the resonator is superradiant, and scales as $N^2$.

The light field emitted into the cavity can interfere with the incident field to form an optical lattice that is sufficiently strong to influence the motion and spatial distribution of a laser cooled atomic gas. In this case, self-organization can set in suddenly as a phase transition above a certain incident pumping threshold \cite{Domokos02,Nagy10,Keeling10,Fernandez-Vidal10}, as observed both for a cold thermal ensemble \cite{Black03,vonCube04} and for a Bose-Einstein condensate \cite{Baumann10,Slama07,Bux11}.

\section{Quantum mechanical expression for the cooperativity parameter}

Having concluded our purely classical treatment of atom-cavity interactions, we now show that our definition of the cooperativity parameter is equivalent to the standard cavity QED definition in terms of the quantum mechanical vacuum Rabi frequency $2g$ \cite{Kimble98,API98,Haroche06}.  There, $g$ is given by the atom's dipole coupling $g=\mu E_v/\hbar$ to the RMS vacuum field $E_v$ at an antinode of a cavity mode at the atomic transition frequency $\omega_0=c k_0$.  The vacuum energy in this mode is
\begin{equation}\label{eq:VacEnergy}
\frac{1}{2}\hbar\omega_0= \epsilon_0 E_v^2 V,
\end{equation} where $V=\int \exp(-2\rho^2/w^2)\sin^2(k_0 z) 2\pi\rho d\rho dz=\pi w^2 L/4$ represents the mode volume.  Thus,
\begin{equation}
g=\mu\sqrt{\frac{\omega_0}{2\epsilon_0\hbar V}}.
\end{equation}

We have already suggested a relation between the vacuum Rabi frequency $2g$ and the normal mode splitting
\begin{equation}\label{Eq:VacuumRabi}
2g_\mathrm{cl} = \sqrt{\eta\kappa\Gamma}
\end{equation}
appearing in the cavity transmission and atomic emission spectra derived in section \ref{Sec:CavityAbsorption}.  That this classically derived normal-mode splitting is indeed identical to the vacuum Rabi frequency in cavity QED can be verified by substituting into Eq. \ref{Eq:VacuumRabi} the cooperativity $\etac = 24 \F /(\pi k_0^2 w^2)$ from Eq. \ref{Eq:EtaCavity}, the cavity linewidth $\kappa = \pi c/(L\F)$, and the atomic excited-state linewidth $\Gamma = k_0^3 |\mu|^2/(3 \pi \epsilon_0 \hbar)$.  One obtains
\begin{equation}
g_\mathrm{cl}=\mu\sqrt{\frac{2\omega_0}{\epsilon_0\hbar\pi w^2 L}}=g.
\end{equation}

Rearranging Eq. \ref{Eq:VacuumRabi} thus gives the standard quantum mechanical expression \cite{Kimble98} for the cooperativity parameter as an interaction-to-decay ratio:
\begin{equation}\label{Eq:etaQM}
\etac=\frac{4g^2}{\kappa\Gamma}.
\end{equation}
Note that this expression for $\etac < 1$ can also readily be interpreted as the cavity-to-free-space scattering ratio, since the rate at which an excited atom emits into the cavity is given by Fermi's Golden Rule as $4g^2/\kappa$.

\section{Conclusion}
We have shown that a variety of fundamental features of the atom-cavity interaction can be described in classical terms, and that the dimensionless cooperativity parameter $\etac$ that scales with the beam area, rather than the beam volume, plays a central role in the classical description. The weak and strong regime can be distinguished by the condition $\etac \lessgtr 1$, which quantum mechanically corresponds to a single-photon Rabi frequency that is small or large compared to the geometric mean of the atomic and cavity linewidths. In the strong-coupling regime even an optical resonator mode that subtends a small solid angle can increase or substantially decrease the emission into free space by the atom, due to the backaction of the cavity field on the atomic dipole.

The classical model is valid at low saturation of atomic transitions, be it due to low beam intensity or large detuning from atomic resonances. The limit of low saturation of the atomic transition exists even if a single cavity photon saturates the atomic transition, i.e. for $2g>\Gamma$ or critical photon number less than one in cavity QED terms. In this case a weak coherent state with less than the critical photon number on average needs to be used to avoid atomic saturation. Then the classical description used here will remain valid.

Most applications of the atom-cavity interaction rely on the narrowband coherent scattering by the atom that can be correctly described in classical terms. The classical model is easily expanded to include the interaction of an atomic ensemble and a cavity mode. In this case the collective cooperativity parameter depends strongly on the ordering of the ensemble.

It is particularly noteworthy that even the strong-coupling regime of cavity QED, giving rise to a normal-mode or ``vacuum Rabi splitting" \cite{Zhu90} can be described in classical terms. One may even ask with \cite{Dowling93} ``How much more classical can you get?", a viewpoint that we cannot completely disagree with.

\section{Acknowledgements}
Over the past years several colleagues have contributed to a deeper, simpler, and more intuitive understanding of the interaction between atoms and cavities, among them Adam Black, Hilton Chan, Yu-ju Lin, Igor Teper, and James Thompson. Dan Stamper-Kurn was the first to point out to one of us (V.V.) that the cavity-to-free-space scattering ratio introduced in the context of cavity cooling \cite{Vuletic00} is the same as the cooperativity parameter in cavity QED. We also acknowledge inspiring discussions with Isaac Chuang, Tilman Esslinger, Mikhail Lukin, Jakob Reichel, and Helmut Ritsch. We thank Paul Berman for critical reading of the manuscript and discussions.

We gratefully acknowledge support by the NSF, DARPA, and ARO.




\bibliography{vv_ref10,vv_ref11_off_ht,vv_ref11_laptop}


\end{document}